\documentclass[fleqn,usenatbib]{mnras}

\usepackage{newtxtext,newtxmath}

\usepackage[T1]{fontenc}
\usepackage{ae,aecompl}

\usepackage{graphicx}	
\usepackage{amsmath}	
\usepackage{amssymb}	
\usepackage[usenames,dvipsnames]{xcolor} 
\usepackage[normalem]{ulem} 

\newcommand{\bluetides}{\mbox{\sc{BlueTides}}}
\newcommand{\ceagle}{\mbox{\sc{C-Eagle}}}
\newcommand{\eagle}{\mbox{\sc{Eagle}}}
\newcommand{\euclid}{\mbox{\it Euclid}}
\newcommand{\hst}{\mbox{\it HST}}
\newcommand{\jwst}{\mbox{\it JWST}}
\newcommand{\flares}{\mbox{\sc Flares}}
\newcommand{\flare}{\mbox{\sc Flare}}
\newcommand{\rst}{\mbox{\it Roman Space Telescope}}
\newcommand{\Msun}{M$_{\odot}$}

\newcommand{\eg}[0]{$\textnormal{e.g. }$}
\newcommand{\ie}[0]{$\textnormal{i.e. }$}
\newcommand{\etc}[0]{$\textnormal{etc}$}

\definecolor{deepchestnut}{rgb}{0.73, 0.31, 0.28}

\defcitealias{lovell2020}{\flares\,I}

\title[FLARES II]{First Light And Reionisation Epoch Simulations\\ (FLARES) II: The Photometric Properties of High-Redshift Galaxies}

\author[Vijayan et al.]{Aswin P. Vijayan$^{1}$\thanks{E-mail: A.Payyoor-Vijayan@sussex.ac.uk},
Christopher C. Lovell$^{2,1}$,
Stephen M. Wilkins$^{1}$,
Peter A. Thomas$^{1}$,
\newauthor
David J. Barnes$^{3}$, 
Dimitrios Irodotou$^{1}$, 
Jussi Kuusisto$^{1}$,
William J. Roper$^{1}$ 
\\
$^{1}$Astronomy Centre, University of Sussex, Falmer, Brighton BN1 9QH, UK\\
$^{2}$Centre for Astrophysics Research, School of Physics, Astronomy $\&$ Mathematics, University of Hertfordshire, \\Hatfield AL10 9AB, UK\\
$^{3}$Department of Physics, Kavli Institute for Astrophysics and Space Research, Massachusetts Institute of Technology,\\Cambridge, MA 02139, USA
}

\date{Accepted XXX. Received YYY; in original form ZZZ}

\pubyear{2020}

\begin{document}
\label{firstpage}
\pagerange{\pageref{firstpage}--\pageref{lastpage}}
\maketitle

\begin{abstract}
We present the photometric properties of galaxies in the First Light and Reionisation Epoch Simulations (\flares). The simulations trace the evolution of galaxies in a range of overdensities through the Epoch of Reionistion (EoR). With a novel weighting scheme we combine these overdensities, extending significantly the dynamic range of observed composite distribution functions compared to periodic simulation boxes. \flares\, predicts a significantly larger number of intrinsically bright galaxies, which can be explained through a simple model linking dust-attenuation to the metal content of the interstellar medium, using a line-of-sight (LOS) extinction model. With this model we present the photometric properties of the \flares\, galaxies for $z \in [5,10]$. We show that the ultraviolet (UV) luminosity function (LF) matches the observations at all redshifts. The function is fit by Schechter and double power-law forms, with the latter being favoured at these redshifts by the \flares\, composite UV LF. We also present predictions for the UV continuum slope as well as the attenuation in the UV. The impact of environment on the UV LF is also explored, with the brightest galaxies forming in the densest environments. We then present the line luminosity and equivalent widths of some prominent nebular emission lines arising from the galaxies, finding rough agreement with available observations. We also look at the relative contribution of obscured and unobscured star formation, finding comparable contributions at these redshifts. 
\end{abstract}

\begin{keywords}
galaxies: general -- galaxies: evolution -- galaxies: formation -- galaxies: high-redshift -- galaxies: photometry 
\end{keywords}



\section{Introduction}\label{sec:intro}
The past few decades have seen tremendous growth in the understanding of galaxy formation and evolution in the first billion years of the Universe after the Big Bang. The first stars and galaxies formed within the first few million years after the big bang. These were the first sources of ionising photons in the Universe, ushering in the Epoch of Reionisation (EoR) by ionising hydrogen \citep[\eg][]{Wilkins2011a,Bouwens2012b,Robertson2013,Robertson2015,Dayal2018}. 

Thanks chiefly to the efforts of the \textit{Hubble Space Telescope} \citep[\hst\,, \eg][]{Beckwith_2006,Bouwens2008,Labbe2010,Robertson2010,Wilkins2010,Bouwens2014a,McLeod2015,Bowler2017,Kawamata_2018} and the \textit{Visible and Infrared Survey Telescope for Astronomy} \citep[\textit{VISTA}, \eg][]{Bowler2014, Stefanon2019,Bowler2020} more than a thousand galaxies have now been identified at $z>5$ with a handful of candidates even identified at $z>10$ \citep[\eg][]{Oesch_2016,Bouwens_2019}. These efforts have also been complemented by \textit{Spitzer} providing rest-frame optical photometry \citep[\eg][]{Ashby_2013,Roberts_Borsani_2016,Bridge_2019} and the \textit{Atacama Large Millimeter/submillimeter Array}  \citep[\textit{ALMA}, \eg][]{Smit2018,Carniani2018,Hashimoto2019} providing rest-frame far-IR and sub-mm photometry and spectroscopy.

With upcoming facilities like the \textit{James Webb Space Telescope}, \euclid\,, and the \textit{Nancy Grace Roman Space Telescope} that can comprehensively study galaxies in the EoR, it is timely to model and predict the properties of these high redshift systems. The \textit{Webb Telescope} will be able to provide better sensitivity and spatial resolution in the near and mid-infrared, providing rest-frame UV-optical imaging and spectroscopy. \euclid\, and \rst\, can do deep and wide surveys adding better statistics to the bright end. The combined efforts of both these observatories can thus provide effective constraints on the bright and rare galaxies in the early Universe. These next generation of surveys would be the test beds to further the theory of galaxy formation and evolution. 

One of the quantities in the EoR where we have extensive observational constraints is the galaxy UV luminosity function, measuring the comoving number density of galaxies as a function of their luminosity across different redshifts. There have been numerous studies \cite[\eg][]{Bouwens_2015a,McLeod2015,Finkelstein2015,Livermore2017,Atek2018,Stefanon2019,Bowler2020} done to quantify this function, providing better understanding of this population. 

Another exciting area which is currently being probed are line luminosities and their equivalent widths. Lyman-$\alpha$ has been primarily used for spectroscopic confirmation of high-redshift galaxies, but becomes increasingly weak at high-redshift due to increasing neutral fraction in the inter-galactic medium (IGM). Rest-frame far-infrared lines are also a useful probe of galaxies in the EoR, serving as diagnostics of the physical and chemical conditions of the inter-stellar medium (ISM) phases. ALMA has had mixed success in detecting the brightest of the far-infrared fine-structure lines like [CII] and [OIII] in the EoR. However, it has detected these lines even in some of the highest redshift galaxies \citep[\eg][]{Hashimoto2018,Harikane2020}. Some works have also looked at rest-frame optical line emission like the the [OIII] and [CIII] doublet \cite[\eg][]{Stark2015,Stark2017,deBarros19_OIIIHbeta}, providing a window into the nature of the ionizing radiation field in these galaxies. These observations have also found extreme equivalent width values in some galaxies. Many of the emission lines in the optical arise from HII regions rather than from photo-dissociation regions (PDRs), making their modelling easier compared to the latter. Most of the existing constraints on galaxy properties in the EoR come from luminosity functions in the UV; this will change with the launch of the \textit{Webb Telescope}, whose onboard instruments will provide access to many of the strong emission lines in the EoR. 

Complementary to this, many theoretical works on simulations of galaxy evolution have been used to study the population of galaxies and their properties in the EoR \cite[\eg][]{Mason2015,Wilkins2017,Ceverino2017,Ma2018,Finlator2018,Yung2019a,Wu2020}. There are various intrinsic physical properties of galaxies, like stellar mass and star formation rate, that are available directly from simulations, which can be compared to that of observed galaxies. These all involve some modelling assumptions based on the star formation history or metallicity of the observed galaxies, which are hard to derive with limited available data on the galaxy at these high redshifts. Another approach is to make predictions from simulations to compare to galaxy observables that suffer from comparatively less modelling biases such as luminosities and line equivalent widths, thus providing insights into the physical processes that take place in these galaxies.  

Semi-Anlytical Models (SAMs), which run on halo merger trees extracted from dark matter only simulations or Extended Press-Schechter methods, have been widely used and very successful in the study of galaxy formation and evolution \cite[\eg][]{henriques2015,Somerville2015,Rodrigues2017,henriques2020}. A number of these studies have been used to make predictions on the observables in the EoR \cite[\eg][]{Clay2015,Mason2015,Poole2016,Lacey2016,Yung2019b,Hutter2020,Dayal2020}. They are powerful tools that can be applied to large cosmological volumes thus probing a large dynamic range of various distribution functions or observables due to their shorter computation times. With each generation of SAMs, there are more detailed physical models being incorporated in them. However they treat galaxies as unresolved objects, modelling various components of galaxy evolution with their integrated properties. Hence, they do not self-consistently evolve
various interactions such as mergers and feedback events, requiring additional steps and approximations to retrieve observables.

In contrast, hydrodynamical simulations of galaxy formation model in greater detail the evolution of dark matter, gas, stars and black holes, allowing for a more detailed exploration of galaxy structure and observed properties. Many state of the art periodic cosmological volumes like \textsc{MassiveBlack} \citep{Di_Matteo_2012}, \textsc{Illustris} \citep{vogelsberger_introducing_2014,Vogelsberger2014,Genel2014,Sijacki2015}, \textsc{MassiveBlack-II} \citep{Khandai2015}, \eagle\, \citep{schaye_eagle_2015, crain_eagle_2015}, \bluetides\, \citep{Feng2016}, \textsc{Mufasa} \citep{dave_mufasa:_2016-1}, \textsc{Cosmic Dawn} \citep{Ocvirk2016}, \textsc{Illustris-TNG} \citep{Naiman2018,Nelson2018,Marinacci2018,Springel2018,Pillepich2018}, \textsc{Simba} \citep{dave_simba:_2019}, \textsc{Cosmic Dawn II} \citep{Ocvirk2020}, \etc\, have been undertaken independently and have been successful in reproducing many of the observables. However, their volumes are too small to replicate many of the current observations of massive galaxies at the bright end, which are born in rare overdensities in the EoR. The enormous computational time to run such large periodic volumes have been a major roadblock from exploring large dynamic ranges with better resolution. 

A successful approach to tackle this limitation has been the use of zoom simulations, whose regions are drawn from less expensive, low-resolution dark matter only simulations, whose box lengths can be in the gigaparsecs. These can be run at higher resolution with additional physics, by generating the initial conditions of the required patch of volume. This approach preserves the large-scale power and the long-range tidal forces by simulating the matter outside the volume of interest at a much lower resolution. For instance, this technique has been successfully employed to re-simulate cluster environments \cite[similar to the works of][\etc]{Bonafede2011,Planelles2014,Pike2014} in the \textsc{C-Eagle} simulations \citep{barnes_cluster-eagle_2017,bahe_hydrangea_2017}, whose regions were selected from a parent dark matter only simulation box of side length 3.2 cGpc \citep{barnes_redshift_2017}. The simulations used the \eagle\, physics model, allowing the model to be used in cluster environments without the need to simulate large periodic boxes. There have also been high resolution zoom simulations that have probed the galaxy properties in the EoR like the stellar mass function or the luminosity function \cite[\eg][]{Ceverino2017,Ma2018} as well as the Lyman-$\alpha$/Lyman-continuum studies \cite[\eg][]{Katz2018} or line emissions \cite[\eg][]{Pallottini2019}. Non zoom, high resolution cosmological simulation \textsc{Sphinx} \citep{Rosdahl2018}, has also been used to study reionisation histories. However they have not necessarily extended the dynamic range that will be probed by the next generation surveys.

The zoom technique can also be applied to get representative samples of the Universe. An example of this, was the GIMIC simulations \citep{crain_galaxies-intergalactic_2009}, which sampled 5 regions of various overdensities from the dark matter only Millennium simulation \citep{springel_simulations_2005} at $z=1.5$. These regions were then re-simulated at a higher resolution with full hydrodynamics. In this case one can produce composite distribution functions by combining the regions using appropriate weights based on their overdensity. This allows for the exploration of the environmental effects of galaxy formation as well as extend the dynamic range of distribution functions without the need to simulate large boxes. Another example is the use of FIRE-2 \citep{Hopkins2018} physics model in \cite{Ma2018}, to re-simulate various halos selected at $z=5$ from dark matter only simulation boxes (largest box used is of side length 43 cMpc) at higher resolution. The re-simulated galaxies are combined with a weighting scheme based on the abundance of the target halos in the Universe, to produce composite distribution functions. 

For the purpose of studying the EoR, we have run a suite of zoom simulations, termed First Light and Reionisation Epoch Simulations, \textsc{Flares}; introduced in \cite{lovell2020} (hereafter \citetalias{lovell2020}), using the \eagle\, \citep{schaye_eagle_2015,crain_eagle_2015} model to re-simulate a wide range of overdensities in the EoR.  \flares\, follows an approach similar to the GIMIC simulations to produce composite distribution functions. 

\citetalias{lovell2020} investigated some of the galaxy properties like the stellar mass function, the star formation rate function and the impact of environment at high redshift. In this, second \flares\, paper, we use the suite of re-simulations to study the photometric properties of the galaxies in the EoR which will be accessible to the upcoming \textit{Webb}, \euclid\,,  \textit{Roman} telescopes. We examine the UV LF, UV continuum slope, attenuation in the UV as well as the effect of environment on the UV LF. We also study the line luminosities and equivalent widths of some of the prominent nebular emission lines. In addition to this we also look at the contribution of the obscured and unobscured star formation rate in the EoR. 

We begin by briefly introducing the simulation suite in Section~\S\ref{sec:simulations} and our modelling of galaxy observables in Section~\S\ref{sec:simulations.SED} and \S\ref{sec:modelling.dust}. In Section~\S\ref{sec:PhotProp} we focus on the derived photometric properties of the simulated galaxies like the UV LF and nebular line emission properties. In \S\ref{sec: sfr} we investigate the fraction of obscured and unobscured star formation rate in the EoR, and present our conclusions in Section~\S\ref{sec:conc}. We assume a Planck year 1 cosmology \cite[$\Omega_{m}$ = 0.307, $\Omega_{\Lambda}$ = 0.693, h = 0.6777;][]{planck_collaboration_2014}. 
\section{The \flare\, Simulations}\label{sec:simulations}

\textsc{Flares} is a suite of zoom simulations targeting regions with a range of overdensities in the Epoch of Reionisation (EoR). 
These regions are drawn from the same $(3.2\; \mathrm{cGpc})^{3}$ dark matter only, parent simulation box used in the \ceagle\, simulations \citep{barnes_redshift_2017}. These regions are then re-simulated until $z=4.67$ with full hydrodynamics using the AGNdT9 configuration of the \eagle\, galaxy formation model, as described in \cite{schaye_eagle_2015,crain_eagle_2015}. The simulations have an identical resolution to the 100 cMpc Eagle Reference simulation box, with a dark matter and an initial gas particle mass of $\mathrm{m_{\text{dm}}} = 9.7\times10^6 \; \mathrm{M_{\odot}}$ and $\mathrm{m_{\textrm{g}}} = 1.8\times10^6 \; \mathrm{M_{\odot}}$ respectively, and has a gravitational softening length of $2.66 \; \mathrm{ckpc}$ at $z\ge2.8$. 

\eagle\,, is a series of cosmological simulations, run with a heavily modified version of \textsc{P-Gadget-3}, which was last described in \cite{springel_simulations_2005}, an N-Body Tree-PM smoothed particle hydrodynamics (SPH) code. The model uses the hydrodynamic solver collectively known as \textsc{Anarchy} \citep[described in][]{schaye_eagle_2015,Schaller2015}, that adopts the pressure-entropy formulation described by \cite{Hopkins2013}, an artificial viscosity switch \citep{cullen_inviscid_2010}, and an artificial conduction switch \citep[\eg][]{price_modelling_2008}. The model includes radiative cooling and photo-heating \citep{Wiersma2009a}, star formation \citep{Schaye2008}, stellar evolution and mass loss \citep{Wiersma2009b}, black hole growth \citep{springel_simulations_2005} and feedback from star formation \citep{DallaVecchia2012} and
AGN \citep{springel_simulations_2005,B_and_S2009,Rosas-Guevara2015}. The subgrid model was calibrated to reproduce the observed $z=0$ galaxy mass function, the mass-size relation for discs, and the gas mass-halo mass relation. The model has also been found to be in good agreement for a number of low-redshift observables not used in the calibration \citep[\eg][]{furlong_evolution_2015,Trayford2015,Lagos2015}. The AGNdT9 configuration produces similar mass functions to the Reference model but better reproduces the hot gas properties of groups and clusters \citep{barnes_cluster-eagle_2017}. It uses a higher value for C$_{\text{visc}}$, a parameter for the effective viscosity of the subgrid accretion, and a higher gas temperature increase from AGN feedback, $\Delta$T. These modifications give less frequent, more energetic AGN outbursts. 

The selection of regions from the parent box is done at $z=4.67$, from which we select 40 spherical regions with a radius of 14 cMpc/h, spanning a wide range of overdensities, ranging from an overdensity value of $\delta=-0.479\to0.970$ (shown in Table A1 of \citetalias{lovell2020}). This redshift selection also automatically ensures that the extreme overdensities are only mildly non-linear, and thus approximately preserves the rank ordering of overdensities at higher redshifts. We have deliberately selected a greater number of extreme overdensity regions (16) to obtain a large sample of the first massive galaxies that are thought to be biased to such regions \citep{chiang_ancient_2013,lovell_characterising_2018}. The range of overdensities allows for better sampling of the density space and explore the impact of environment on galaxy formation and evolution. 

In order to obtain a representative sample of the Universe, these regions are combined using appropriate weightings, with the very overdense and underdense regions contributing the least to the total weight, thus compensating for any oversampling of the overdense regions. With this weighting technique, we are able to probe a bigger volume without drastically lowering the resolution. For a more detailed description of the simulation and weighting method we refer the readers to \citetalias{lovell2020}.

\subsection{Galaxy Identification}\label{sim.galident}

\begin{figure*}
	\centering
	\includegraphics[width=0.9\textwidth]{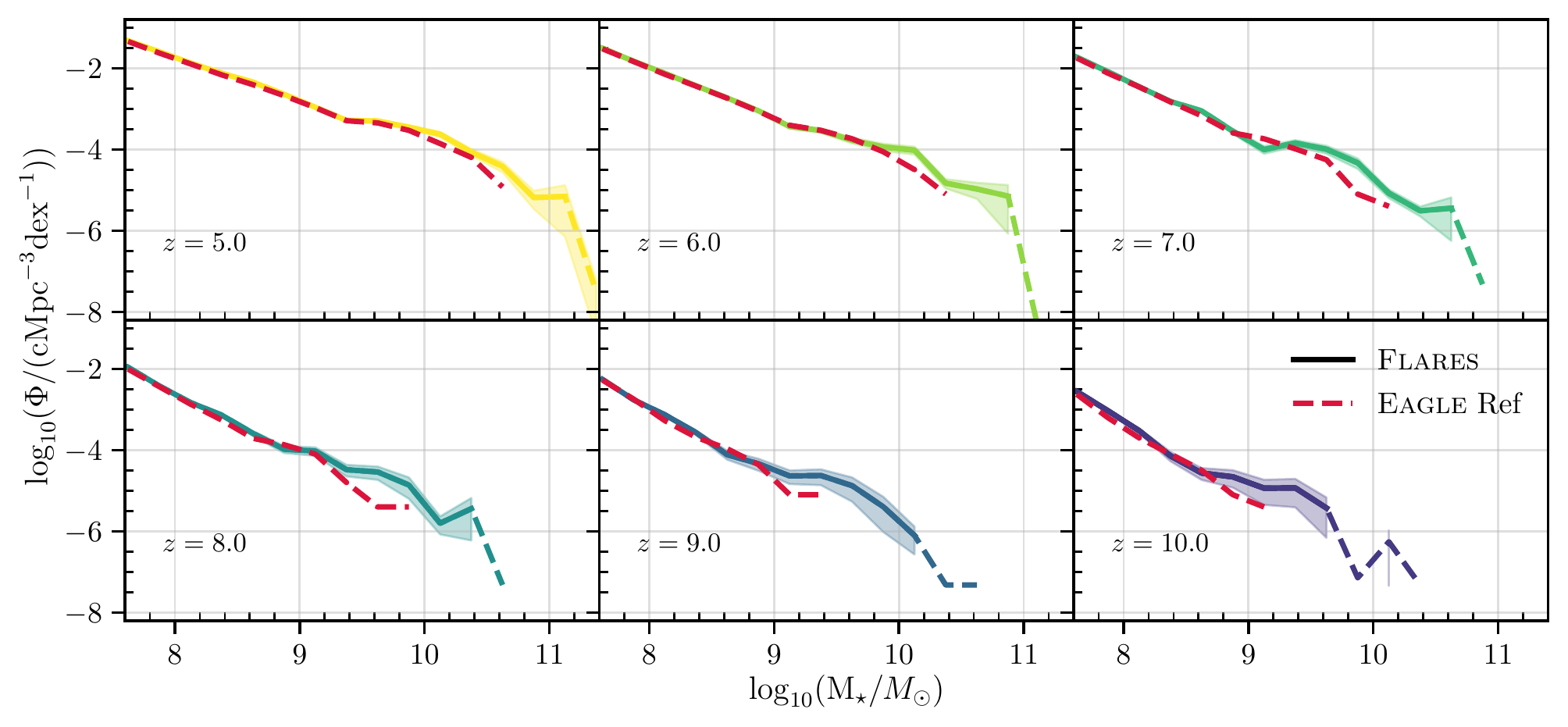}
	\caption{\flares\, composite galaxy stellar mass function (black solid, dashed for bins with less than 5 galaxies) for z $\in$ [5,10]. Shaded regions denote the Poisson $1\sigma$ uncertainties for	each bin from the simulated number counts for the \flares\, galaxies. For comparison the GSMF from the 100 cMpc \eagle\, Reference simulation box is shown in red. \label{fig: GSMF}} 
\end{figure*}
\begin{figure}
	\centering
	\includegraphics[width=0.9\columnwidth]{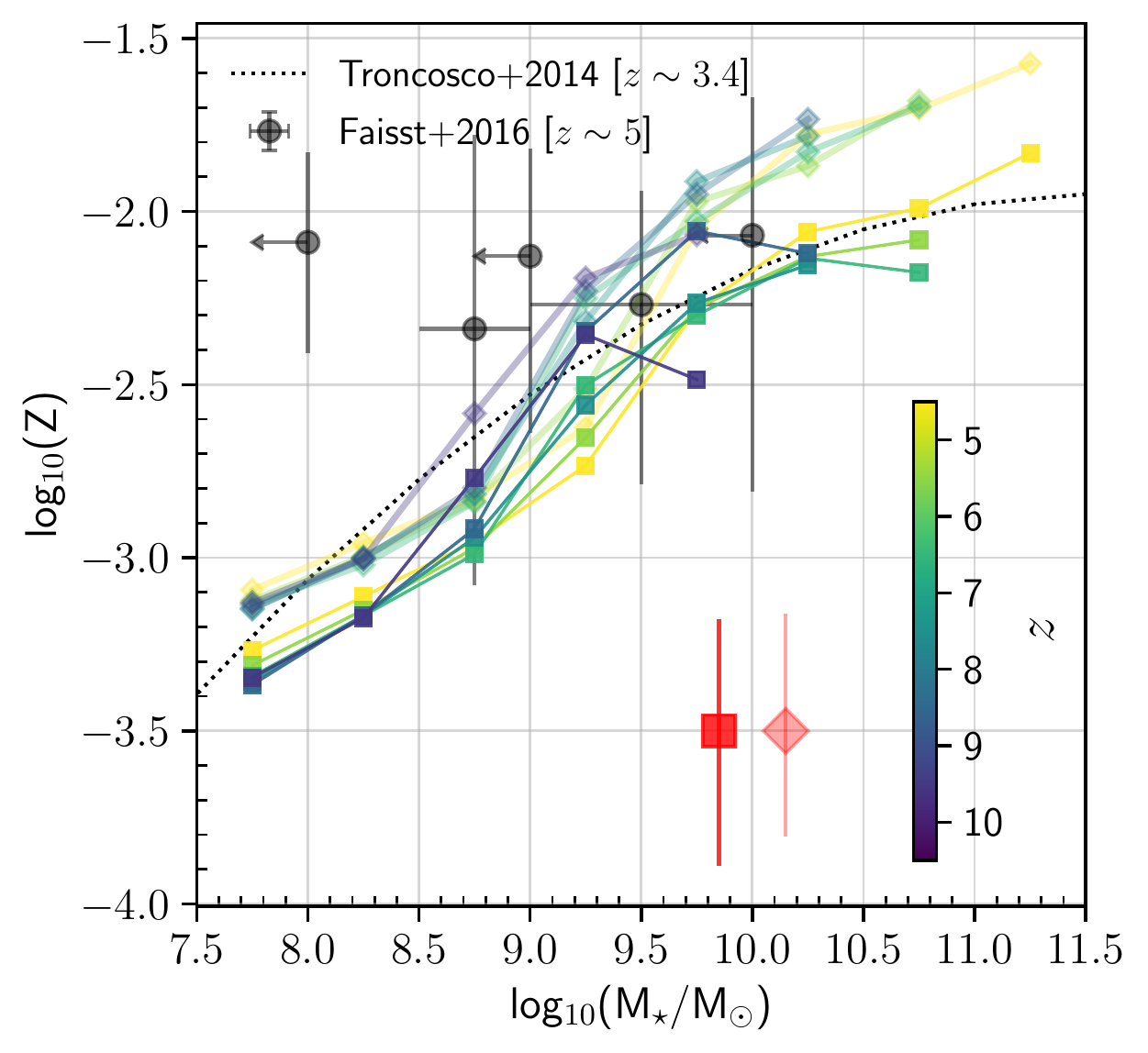}
	\caption{Mass weighted metallicities of the gas (darker square points) and stars (lighter diamond points) of the \flares\, galaxies at $z\in[5,10]$. Only the weighted median of the bins containing more than 5 galaxies are shown, with the maximum of the 16$^{\mathrm{th}}$ and 84$^{\mathrm{th}}$ percentile spread in the bins of the two data shown in red. The observational constraints on the gas-phase metallicity from \protect\cite{Troncoso2014} at $z\sim3.4$ and \protect\cite{Faisst2016} at $z\sim5$ are shown. Observational measurements of the stellar mass assume a \protect\cite{chabrier_galactic_2003} initial mass function with metallicities converted to a mass-fraction assuming $12+\mathrm{log}_{10}(\mathrm{O/H})_{\odot}=8.69$ and Z$_{\odot}=0.02$.\label{fig: metallicity}} 
\end{figure}
Galaxies in \flares\,, similar to the standard \eagle\, are identified with the \textsc{Subfind} \citep{springel_populating_2001,Dolag2009} algorithm, which runs on bound groups found from via the Friends-Of-Friends \citep[FOF,][]{davis_evolution_1985} algorithm. The stellar masses are defined using star particles within a 30 pkpc aperture centred on the most bound particle of the self-bound substructures. In this work, we concentrate on a broader definition of a galaxy with respect to \citetalias{lovell2020}, where only galaxies with a stellar mass $\ge 10^{8}$M$_{\odot}$ were considered in the analysis. Here we focus on objects with a combined total of more than 100 gas and star particles. This extends the stellar mass function down to $\sim$ 10$^{7.5}$\Msun\, at $z=5$.

\flares\, has more than $\sim$20 times the number of galaxies with a mass greater than 10$^{10}$ \Msun\ at $z=5$ compared to the \eagle\, reference volume \citep{schaye_eagle_2015} \citepalias[see Figure~5 in][]{lovell2020}. In Figure~\ref{fig: GSMF}, we compare the galaxy stellar mass function of the galaxies in \flares\, and the 100 cMpc \eagle\, Reference simulation box. It can be seen that \flares\, extends the range by at least an order of magnitude at the high-mass end compared to \eagle\,. 

\subsection{Metal Content}\label{sec:simulations.metals}

Stellar evolution enriches galaxies with metals. This is governed by the rate at which stars are formed and the various mass loss events associated with their evolution (\eg\, stellar winds, supernova explosion). The next generation of stars form from this enriched gas and evolve, continuing the cycle of metal enrichment in the galaxy. We show this process in Figure~\ref{fig: metallicity}, where the evolution of the mass-weighted stellar and gas-phase metallicities are plotted as a function of galaxy stellar mass. The metallicity of galaxies generally increases with stellar mass. There is little evolution in the metallicity across redshifts, but a strong evolution with stellar mass by approximately an order of magnitude increase from the lowest to the highest stellar mass bin. The normalisation, as well as the trend in the metallicity with stellar mass, is similar to observed gas-phase metallicity seen in \cite{Troncoso2014} at $z\sim3.4$, obtained using optical strong line diagnostics with the R$_{23}$ parameter \citep[for a summary see][]{Kewley2008}. A similar normalisation of the relation at higher metallicities is seen at $z\sim5$ in \cite{Faisst2016} using strong optical emission lines. It should be noted that the uncertainties on the observed metallicities is very large, due to the difficulty in measuring the value at $z\ge5$. Observations from the upcoming \jwst\, will be able to put tighter constraints in the high-redshift regime.

\subsection{Spectral Energy Distribution Modelling}\label{sec:simulations.SED}

In this section, we detail the spectral energy distribution (SED) modelling of each galaxy. In this work, we model only the emission from stars (including reprocessing by gas and dust) and defer the treatment of accretion on to super-massive black holes to a future work. We broadly follow the approach implemented by \cite{Wilkins2016a,Wilkins2017,Wilkins2018,Wilkins2020} albeit with modifications to the dust modelling as described in \S\ref{sec:modelling.dust}.

\subsubsection{Stellar Emission}

We begin by modelling the pure stellar emission produced by each galaxy. To do this we associate each star particle with a stellar SED according to its age and metallicity (\ie\, a simple stellar population or SSP). Throughout this work we utilise v2.2.1 of the Binary Population and Spectral Synthesis (BPASS) stellar population synthesis (SPS) models \citep{BPASS2.2.1} and assume a Chabrier initial mass function (IMF) throughout \citep{chabrier_galactic_2003}. As explored in \cite{Wilkins2016a,Wilkins2017,Wilkins2018,Wilkins2020} the choice of SPS and IMF can have a large effect on resulting broadband luminosities and emission line quantities.

\subsubsection{Nebular Emission}

Young stellar populations produce significant Lyman-continuum (LyC) emission. To account for the reprocessing of these photons by surrounding gas we associate each young ($t<10$ Myr) star particle with a surrounding H{\sc ii} region (or birth cloud) powered by its LyC emission. To calculate the nebular emission we follow the approach detailed in \cite{Wilkins2020}. In short, the pure stellar spectrum of each star particle is input to the {\sc cloudy} \citep{Cloudy17.02} photo-ionisation code. The metallicity of the associated H{\sc ii} is assumed to be identical to the star particle, and we adopt the same dust depletion factors and relative abundances as \cite{Gutkin2016}. We assume a reference ionisation parameter (defined at $t=1$ Myr and $Z=0.02$) of $\log_{10}U_{S,{\rm ref}}=-2$, a hydrogen density of $\log_{10}(n_{H}/{\rm cm^{-3}})=2.5$, and adopt {\sc cloudy}'s default implementation of Orion-type graphite and silicate grains.

\subsection{Dust Attenuation}\label{sec:modelling.dust}

One of the most important ingredients in generating mock observations involves modelling the attenuation by dust. It has a major impact on the observed properties of galaxies, with almost 30$\%$ of all photons in the Universe having been reprocessed by dust grains at some point in their lifetime \citep{Bernstein2002}. There have been a few studies that have incorporated dust creation and destruction self-consistently into hydrodynamical simulations \citep[\eg][]{Aoyama2017,McKinnon2016a,Gjergo2018,Li2019,Graziani2020}. They have found mixed success in matching many of the observed galaxy properties like the dust-to-stellar mass ratio, the dust-to-gas ratio or the dust-to-metal ratio. Many of these simulations also have information on the grain sizes or the contribution of different dust species to the total dust mass. This additional information can eliminate some of the post-processing assumptions involved in deriving observed properties \citep[\eg][]{Hou2017,McKinnon2018,Kannan2019,Hirashita2020}. However they also involve additional subgrid recipes which are poorly understood, and can get computationally intensive depending on the modelling techniques. A simple alternative is to model the effect of dust based on the properties of the existing stars and gas particles in the simulation. This is usually done by using the metallicity information of the ISM to build a model to attenuate the stellar spectra. They still incorporate information on the spatial distribution of dust and are therefore more detailed than a simple screen model. 

In this work, for estimating the dust attenuation, each star particle is treated as a point in space with it's emitted light reaching the observer through the intervening gas particles. We fix the viewing angle to be along the z-axis. For the purpose of this study we link the metal column density ($\Sigma\,(x,y)$) integrated along the LOS (z-axis in this case) to the dust optical depth in the V-band (550nm) due to the intervening ISM $\tau_{\textrm{ISM,V}}(x,y)$, with a similar approach as in \cite{Wilkins2017}. This relation can be expressed as
\begin{equation}\label{eq: tau}
\tau_{\textrm{ISM,V}}(x,y) = \mathrm{DTM}\,\kappa_{\textrm{ISM}}\,\Sigma\,(x,y)\:,
\end{equation} 
where DTM is the dust-to-metal ratio of the galaxy and $\kappa_{\textrm{ISM}}$ is a normalisation parameter which we have chosen to match the rest-frame far-UV (1500\AA) luminosity function to the observed UV luminosity function from \cite{Bouwens_2015a} at $z=5$. The DTM value of a given galaxy comes from the fitting function presented in \cite{Vijayan2019} (Equation 15 in that work), which is a function of the mass-weighted stellar age and the gas-phase metallicity. This allows for a varying DTM ratio across different galaxies as well as evolution across redshift as seen in observational works \citep[\eg][]{DeVis2019}, depending on their evolutionary stage. This provides a single DTM value per galaxy, assuming no spatial variation. $\kappa_{\textrm{ISM}}$ acts as a proxy for the properties of dust, such as the average grain size, shape, and composition. In a companion work, we will explore the impact of a range of different modelling approaches.

\begin{figure}
	\centering
	\includegraphics[width=0.7\columnwidth]{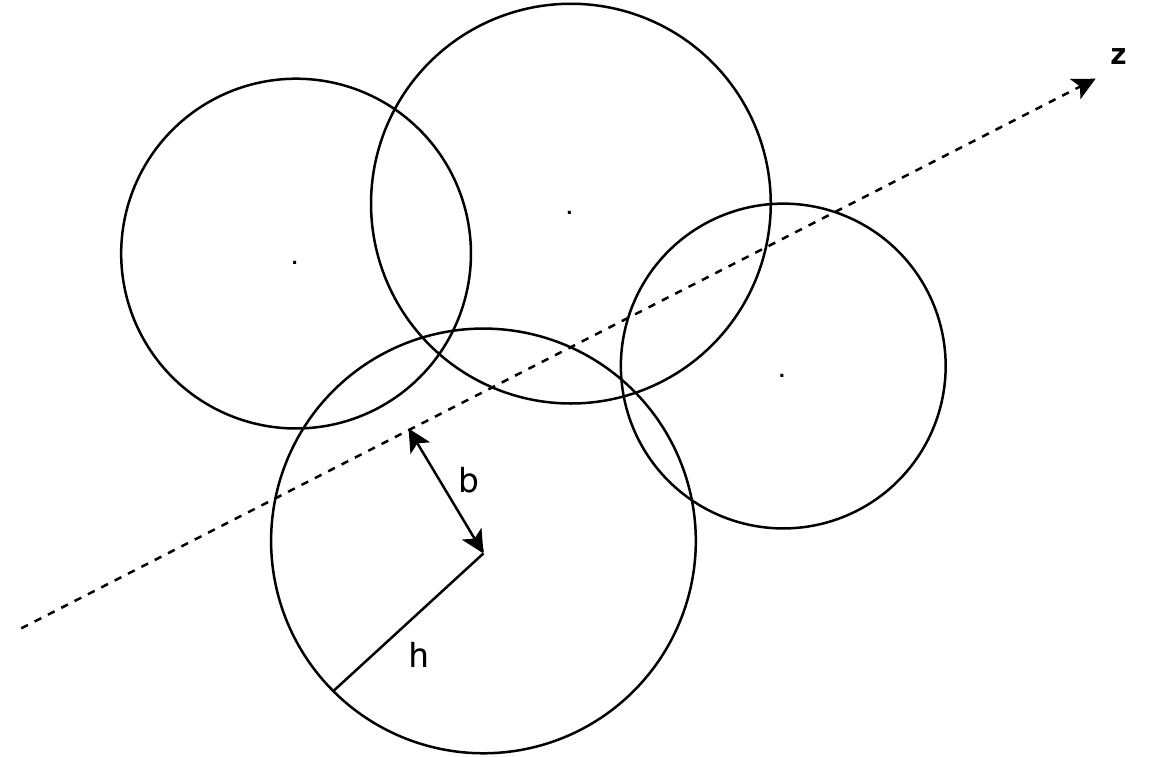}
	\caption{Line of sight tracing of the SPH density field,  with the circles representative of SPH particles. h and b denote the smoothing length of the corresponding gas particle and the impact parameter to the LOS ray respectively. \label{fig: sph_ray}}
\end{figure}

$\Sigma\,(x,y)$ is obtained by integrating the density field of particles along the z-axis with the smoothing kernel of the SPH particle. \textsc{Flares} uses the same flavour of SPH used by \textsc{Eagle}, \textsc{Anarchy} \citep[see][for more details]{Schaller2015}. The kernel function can be expressed as follows:
\begin{equation}\label{eq: kernel}
W(r,h) = \dfrac{21}{2\pi\,h^3} 
\begin{cases}
(1-\frac{r}{h})^4 (1+4\frac{r}{h})& \text{if } 0 \leq r\leq h\\
0  & \text{if }r>h\:,
\end{cases}
\end{equation}
where $h$ is the smoothing length of the corresponding particle and $r$ is the distance from the centre of the particle. The smoothed density line integral across a particular particle can be calculated by using the impact parameter, $b$ which is calculated from the centre of the particle (illustrated in Figure~\ref{fig: sph_ray}). Using the impact parameter of every gas particle in front of the selected stellar particle, the LOS metal column density can be calculated as follows:
\begin{equation}\label{eq: Zlos}
\Sigma\,(x,y) = 2 \sum_i Z_i m_i \int_{0}^{\sqrt{h_i^2-b_i^2}}W(r, h_i)dz\,;\, r^2 = b_i^2 + z^2\:,
\end{equation} 
where the index $i$ denotes gas particles along the LOS, with $Z$ and $m$ the metallicity and mass of the particle respectively. To simplify this calculation, impact parameters can be normalised with the smoothing length, and thus generate pre-computed values of the LOS metal density which can be readily used to compute the density for arbitrary values of smoothing length and impact parameters.

Other than the dust extinction along the LOS, there is an additional component of dust that affects young stellar populations that are still embedded in their birth cloud. Effect of the birth cloud attenuation in our galaxies is a phenomenon that happens below the resolution scale, since stellar clusters form on sub-kpc scales. The birth cloud dust optical depth in the V-band for our model can be expressed in a similar manner to equation \ref{eq: tau} as
\begin{equation}
	\tau_{\textrm{BC,V}}(x,y) = 
	\begin{cases}
	\kappa_{\textrm{BC}} (\textrm{Z}/0.01)\, & \text{t} \leq 10^7\textrm{yr}\\
	0\, & \text{t} > 10^7\textrm{yr}\:,
	\end{cases}
\end{equation}  
where $\kappa_{\textrm{BC}}$ just like $\kappa_{\textrm{ISM}}$, is a normalisation factor, which also encapsulate the dust-to-metal ratio in the stellar birth clouds. This implies that we assume a constant dust-to-metal ratio in birth clouds for all galaxies. Here, $Z$ is the metallicity of the stellar particle with age less than 10$^7$ yr, following the assumption from \cite{Charlot_and_Fall2003} that birth clouds disperse on these timescales. Hence, only young stellar particles are affected by this additional attenuation. With these parameters the optical depth in the V-band is linked to other wavelengths using a simple simple power-law relation
\begin{equation}\label{tau_lambda}
	\tau_{\lambda} = (\tau_{\textrm{ISM}} + \tau_{\textrm{BC}}) \times\,(\lambda/550\textrm{nm})^{-1}\:.
\end{equation}
This functional form yields an extinction curve flatter in the UV than the Small Magellanic Cloud curve \citep{Pei1992}, but not as flat as the \cite{Calzetti2000} curve. 

As discussed earlier there are two free parameters in our model, $\kappa_{\textrm{ISM}}$ that links the optical depth in the ISM to the LOS metal surface density and $\kappa_{\textrm{BC}}$ linking the stellar particle metallicity to the optical depth due to the presence of a birth cloud in young stellar populations. To obtain the values for these parameters we do a simple grid search approach. We make an array of candidate $\kappa_{\textrm{BC}}$ values in the closed range [0.001, 2.]. For each $\kappa_{\textrm{BC}}$, we generate the UV LF for a grid of $\kappa_{\textrm{ISM}}$ values in the range (0, 1] at $z=5$. These are then compared to the \cite{Bouwens_2015a} UV LF at $z=5$ using a simple chi-square analysis to obtain the corresponding value for $\kappa_{\textrm{ISM}}$ (only M$_{\textrm{UV}}<-18$ is used). We then generate the corresponding UV-continuum slope ($\beta$) as well as the [OIII]$\lambda$4959,5007 and H$\beta$ line luminosity and equivalent widths (EW) for a given combination of ($\kappa_{\textrm{BC}}$, $\kappa_{\textrm{ISM}}$). The combination of ($\kappa_{\textrm{BC}}$, $\kappa_{\textrm{ISM}}$) value that best matches the M$_{\textrm{UV}}-\beta$ observations from \cite{Bouwens2012b,Bouwens2014a} at $z=5$ (Figure~\ref{fig: beta diff kappa}) and the [OIII]$\lambda$4959,5007 + H$\beta$ line luminosity and EW relations versus UV luminosity and stellar mass at $z=8$ from \cite{deBarros19_OIIIHbeta} (Figure~\ref{fig: line lum diff kappa}) is chosen as our default model. This process leads a value of $\kappa_{\textrm{BC}}=1$ and $\kappa_{\textrm{ISM}}=0.0795$, which is used for all redshifts considered in this study. A higher value for $\kappa_{\textrm{BC}}$ is favoured to get better agreement with the $\beta$ observations while the line luminosity and EW relations prefer a lower value. Hence the chosen value of $\kappa_{\textrm{BC}}$ is a way to incorporate the effect of both these observations. Future measurements in this observational space from current and upcoming telescopes, would help to further tighten our constraints on this value. The parameter search is explained further in Appendix \ref{app:calibration}. We would also like to remind the reader that by using fixed choice of these parameters, we assume there is no evolution in the general properties of the dust grains in galaxies such as the average grain size, shape, and composition.

We also show in Appendix \ref{app:extinction_curves} how some of the observables presented in the next sections change on using different extinction curves available from literature. 

\section{Photometric Properties}\label{sec:PhotProp}
\subsection{UV Luminosity Function}\label{sec:PhotProp.UVLF}

\begin{figure*}
	\centering
	\includegraphics[width=0.95\textwidth]{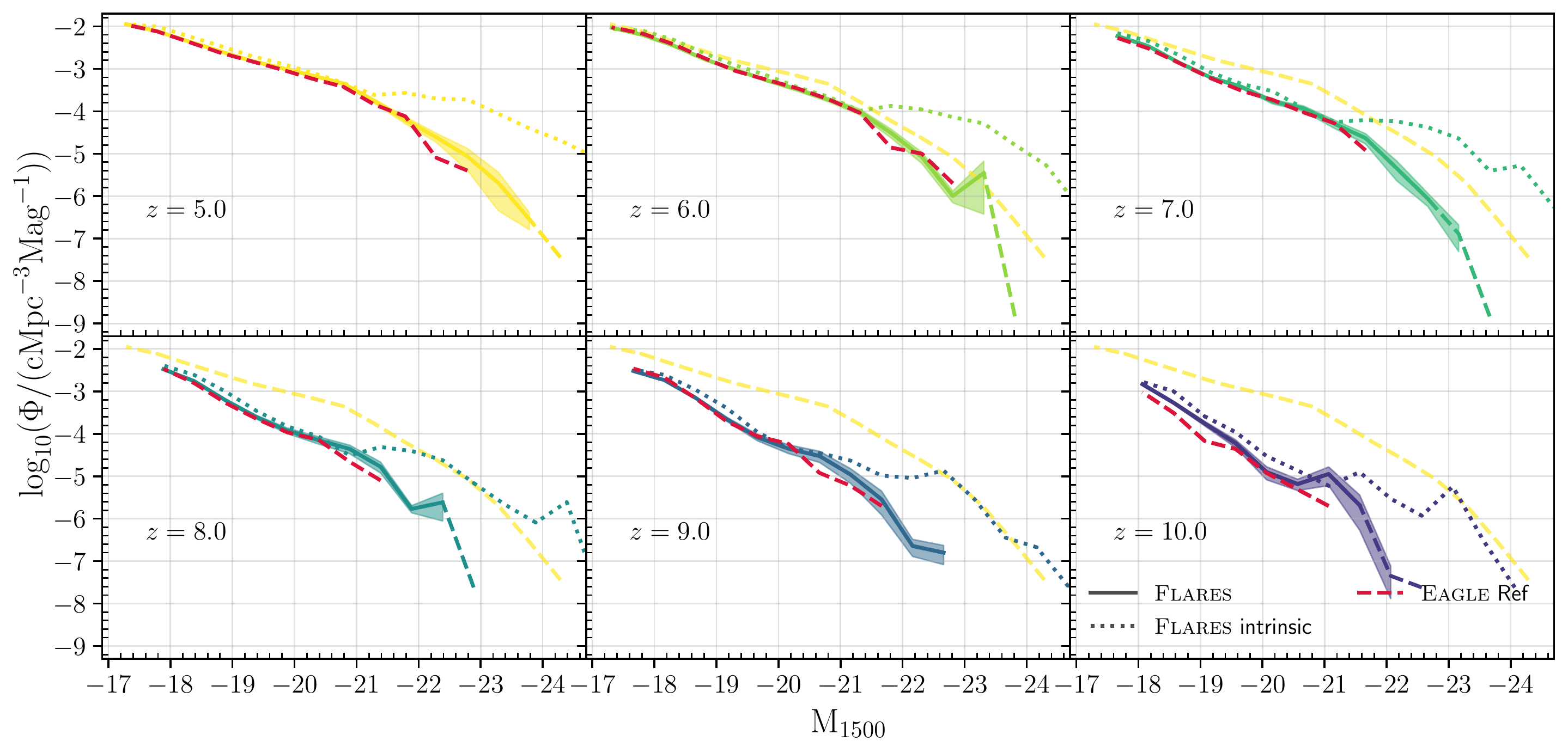}
	\caption{\flares\, composite intrinsic (dotted) and dust attenuated (solid, dashed for bins with less than 5 galaxies) UV LF for  galaxies in z $\in$ [5,10]. Shaded region denote the Poisson $1\sigma$ uncertainties for each bin from the simulated number counts for the dust attenuated UV LF. For comparison the dust attenuated UV LF from the \textsc{Eagle} Reference volume is plotted in red. We also plot the $z=5$ dust attenuated UV LF (dashed line) alongside other redshifts to aid comparison. \label{fig: UVLF}} 
\end{figure*}
The UV LF evolution of high-redshift galaxies is a parameter space where there are numerous observational studies \cite[\eg][]{Bunker2004,Bouwens2006,Wilkins2011a,Bouwens_2015a,Finkelstein2015}. We begin by calculating the rest-frame UV LF of the \flares\, galaxies. 

\subsubsection{LF creation}\label{sec: UVLF.LF_creat}
Unlike cosmological box simulations, the re-simulation strategy of \flares\, means that the creation of the luminosity function (or stellar mass function) is not straightforward. The contribution from any of our re-simulated region needs to weighted by the appropriate weight for that region. This can be explained as follows
\begin{gather}\label{eq: LF weights}
	\mathrm{LF}_i = \Sigma_{j} w_j\, N_{ij}/V\,,
\end{gather}
where LF$_{i}$ is the galaxy number density in bin `$i$', $w_{j}$ is the weight associated with the region `j', $N_{ij}$ is the number of galaxies associated with region `$j$' in bin `$i$' and V is the volume of a single region. Similarly the poisson error associated with a luminosity bin, LFerr$_i$ can be expressed as 
\begin{gather}\label{eq: LF error}
	\mathrm{LFerr}_i = \sqrt{\Sigma_{j} \bigg(w_j\, \sqrt{N_{ij}}\,\bigg)^2}\,/V,
\end{gather}
The weighting scheme of \flares\, has been explained in detail in \S 2.3 in \citetalias{lovell2020}; we refer the reader there for more details.

As described in \S\ref{sim.galident}, we concentrate on a broader definition of a galaxy focusing on only those objects with a combined total of more than 100 gas and star particles, extending the stellar mass function to $\sim$ 10$^{7.5}$\Msun\, at $z=5$. For the luminosity function we set the low brightness cut-off for the selected galaxies to be the 97th percentile of the magnitude computed for 100 gas and star particles, allowing us to probe down to $\sim$-17 in FUV rest-frame magnitude at $z=5$. This also means that most of our galaxies have many more than 100 gas and star particles. 

\subsubsection{Luminosity Functions}\label{sec: PhotProp.UVLF.LF}
We plot the dust-attenuated (as described in \S\ref{sec:modelling.dust}) UV LF in Figure~\ref{fig: UVLF} (solid line) along with the intrinsic LF (dashed line). Here the plotted data for \flares\, are in bins of width 0.5 magnitudes, with their 1$\sigma$ Poisson scatter. Also plotted is the UV LF of the 100 cMpc \eagle\, Reference simulation box. The luminosity function is extended to brighter galaxies by 2 magnitudes or more at all redshifts, with the Reference volume failing to probe the bright end of the UV LF. It is evident that at the faint-end the simulations agree. The bin centre and the number density per magnitude for the \flares\, galaxies are provided in Appendix~\ref{app: lum_func} as Table~\ref{tab: LF values}. 

The number density of bright galaxies (M$_{1500}\le-20$) increases by $\sim2$ orders of magnitude going from $5\to10$ in redshift, indicating the rapid assembly of stars in galaxies through time. It can also be seen that the observed LF is slightly lower than the intrinsic LF at luminosities fainter than $\sim-20$. The reason for this is the implementation of a birth cloud component for young stellar populations. Studies exploring the impact of birth cloud attenuation have shown that this can reduce the luminosities by $\sim\,0.3$ dex for galaxies in the local Universe \citep[\eg][]{Trayford2017}. Since the surface density of metals in the faint galaxies is insufficient to produce significant attenuation in the ISM, the choice of birth cloud component is most pronounced in this regime. While in the case of the bright end, the main contribution is from the dust attenuation in the ISM. 

It is important to take note that both these regimes can be affected by the choice of initial mass function, the SPS model \citep[see][]{Wilkins2016a} and the attenuation law. We also do not take into account the contribution of accretion on to super-massive black holes (SMBH) which is expected to dominate over the contribution of star formation at the extreme bright end \cite[M$_{\textrm{UV}}\lesssim-23$ Magnitude at $z\sim6$, \eg][]{Glikman2011,Giallongo2015,Ono2018}. To give an estimate on the contribution of SMBH to the galaxy luminosity, we perform a simple analysis. The intrinsic bolometric luminosity of the galaxy is compared to the SMBH bolometric luminosity, calculated using
\begin{gather}\label{eq: BH lum}
	\mathrm{L}_{\mathrm{BH, bol}}=\eta\,\frac{\mathrm{dM}_{\bullet}}{\mathrm{dt}}c^2,
\end{gather}	 
where dM$_{\bullet}$/dt is the accretion rate and $\eta$ is the efficiency, assumed to be 0.1. From this analysis we estimate that the fraction of galaxies where the SMBH bolometric luminosity contributes more than $10\%$ to the total luminosity (intrinsic + SMBH) to be negligible at M$_{1500}>-20$. Below this, the fraction rises to a mean value of $\sim25\%$, with a mean contribution of $\sim15\%$ at $z=5$. However, at $z=10$, $\sim40\%$ of galaxies (with M$_{1500}>-20$) host a SMBH that contributes more than $10\%$ to the total bolometric luminosity, with a mean contribution of $\sim30\%$. We remind the reader that these are the bolometric fractions and thus the contribution to the UV can vary widely depending on the obscured nature of the SMBH. The detailed modelling of SMBH luminosities is the focus of a work in preparation.
\begin{figure}
	\centering
	\includegraphics[width=0.97\columnwidth]{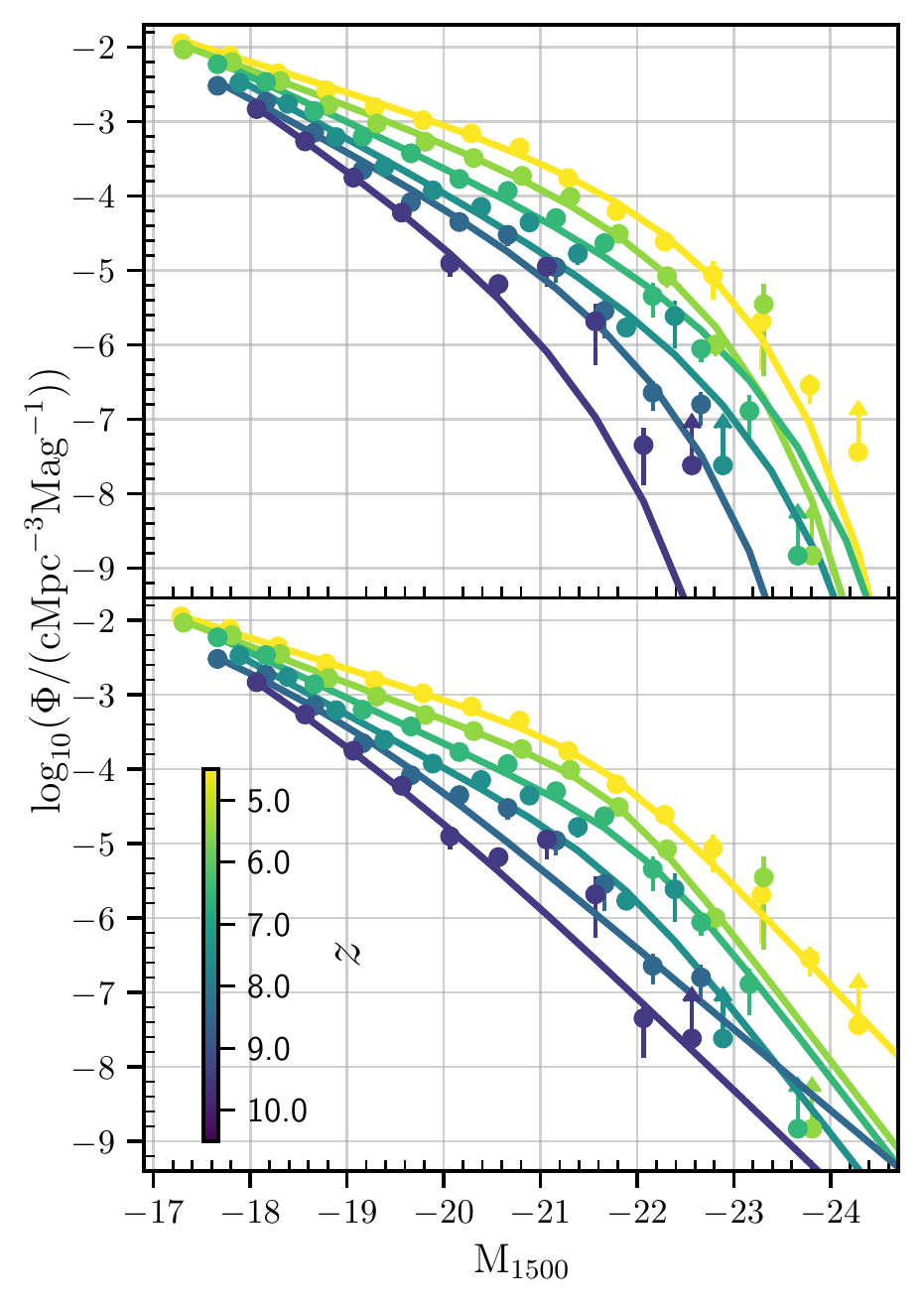}
	\caption{Schechter (top) and double power-law (bottom) fits to the \flares\, UV LF are plotted as solid lines, while the data is shown as points with 1-$\sigma$ Poisson errors. Bins containing single galaxies are indicated by lower limits. \label{fig: LF fit}} 
\end{figure} 
\begin{figure}
	\centering
	\includegraphics[width=0.9\columnwidth]{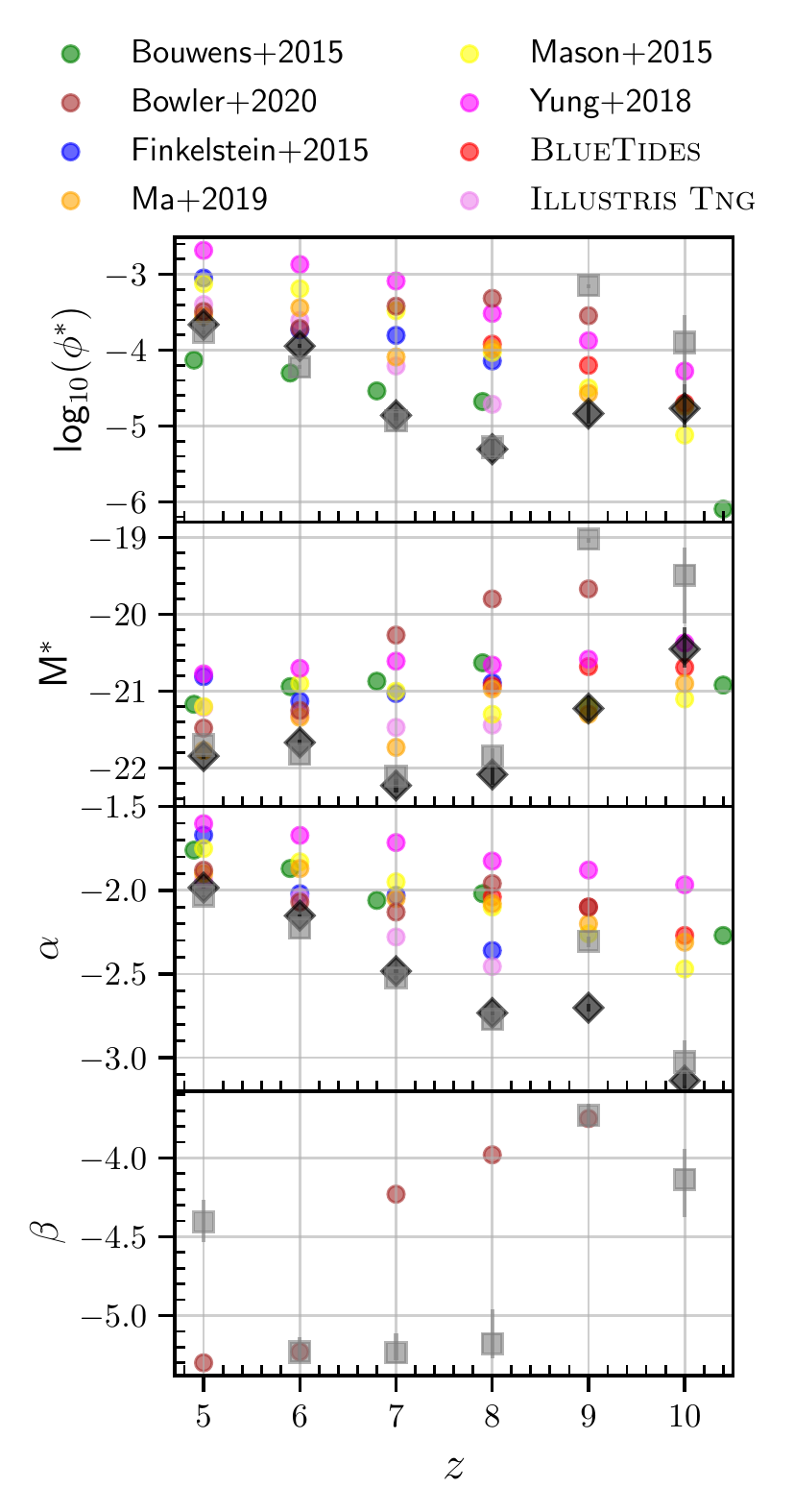}
	\caption{Evolution of the parameters of Schechter (black diamonds) and double power-law (grey squares) fits to the \flares\, UV LF. The quoted error bars show the $16^{\mathrm{th}}-84^{\mathrm{th}}$ percentile uncertainty obtained from the fit posteriors (see Appendix~\ref{app: lum_func} for details) Also plotted are the evolution of the Schechter fit parameters from \bluetides\,\protect\citep{Wilkins2017}, \protect\cite{Bouwens_2015a,Mason2015,Finkelstein2015,Ma2019,yung_semi-analytic_2019}; \textsc{Illustris-Tng} \protect\citep[Model-C from][]{Vogelsberger2020} as well as the double power-law fit parameters from \protect\cite{Bowler2020}. \label{fig: fit param evo}} 
\end{figure}

A Schechter function \citep{schechter_analytic_1976} can be used to describe the UV LF \citep[\eg][]{Bouwens_2015a,Finkelstein2015}, characterized by a power law at the faint end with slope $\alpha$, with an exponential cutoff at the bright end at a characteristic magnitude $M^*$, with the parameter $\phi^*$ setting the normalisation of this function. The number density at a given magnitude is then given by
\begin{equation}\label{eq: schechter}
\phi(M) = 0.4\,\text{ln}\,10\,\phi^{*}\,10^{-0.4(M-M^*)(\alpha+1)}\,e^{-10^{-0.4(M-M^*)}}\:.
\end{equation}
We calculate the Schechter function parameters of our LFs (see Appendix \ref{app: lum_func} for more details of the fitting). The Schechter fits to the UV LF of \flares\, galaxies are shown in Figure~\ref{fig: LF fit} (top panel). We find that the function provides a good fit to the shape of the overall UV LF. The best-fitting Schechter parameters to the UV LF are shown in Table~\ref{tab: fit params}. 

There have also been studies that suggest a double power-law can be used to describe the shape of the UV LF at higher redshifts \citep[\eg][]{Bowler2014}. We describe the parameterization for a double power-law as follows
\begin{equation}\label{eq: DPL}
	\phi(M)  = \frac{\phi^*}{10^{0.4(M-M^*)(\alpha+1)} + 10^{0.4(M-M^*)(\beta+1)}}\:,
\end{equation}
where $\alpha$ and $\beta$ are the faint-end and bright-end slopes,
respectively, $M^*$ is the characteristic magnitude between these two power-law regimes, and $\phi^*$ is the normalisation. The double power-law fit to the binned luminosities is shown in Figure~\ref{fig: LF fit} (bottom panel). The best-fitting double power-law parameters to the UV LF are also shown in Table~\ref{tab: fit params}. It can be seen that this also provides a good fit to the UV LF even though, like the Schechter fit, this parameter form fails to capture the increase in number density around the knee at $z>8$. 

We have already shown in \citetalias{lovell2020} that the galaxy stellar mass function in \flares\, can be described by a double Schechter form. It can be seen in Figure~\ref{fig: UVLF} that the intrinsic UV LF also has a double Schechter shape, but the observed UV LF does not. It lies much closer to a Schechter or a double power-law shape depending on the redshift. This can be explained by dust attenuation suppressing the intrinsically bright galaxies at the knee and beyond. Also shown is the evolution of the parameters of the Schechter and double power-law fits with redshift in Figure~\ref{fig: fit param evo}. We see that for both the fit functions, the value of $M^*$ and $\alpha$ are similar across redshift, with the values generally increasing with increasing redshift for $M^*$ and vice versa for $\alpha$. The Schechter function shows a smooth evolution in all the parameters while in the case of the double-power law there is a sharp upturn in the parameters $\phi^*$, M$^*$ and $\beta$. For the purposes of the fitting (also see Appendix~\ref{app: lum_func}),  $\beta$ was restricted to a lower limit of -5.3, due to the \flares\, LF failing to constrain that parameter. The flatenning at $z\sim7$ can be attributed to this restriction. However, the jump in the parameter space is a consequence of the strong evolution at the bright-end from rapid build up of dust. A similar jump is also seen in the double power-law `$\beta$' parameter presented in \cite{Bowler2020}, albeit at $z=6\to7$.

We compare the performance of the two functional forms across redshifts by computing the Bayesian Information Criterion \citep[BIC, see][and references therein for further details; also see Appendix~\ref{app: lum_func}]{Schwarz1978,Liddle2007} for the best-fit parameters. 
A model with a lower BIC is preferred. For this purpose we give the difference between the BIC values of the double power-law from the Schechter best-fit values, which is also quoted in Table~\ref{tab: fit params}. As can be seen a double power-law function is a much better fit to the UV LF of the \flares\, galaxies at all redshifts, except at $z=10$, where the BIC values are comparable. This could simply be due to the lack of brighter galaxies after the estimated knee of the functions. There are a few explanations in the literature for the emergence of a double power-law shape to the luminosity function at high redshifts. Some studies \citep[\eg][]{Bowler2014,Bowler2020} have suggested that this is due to a lack of evolution in the bright end of the galaxy luminosity function because of the deficit of quenched galaxies at these redshifts. The bright end is very dependent on the dust content as well as star formation of the galaxies, and thus also provides constraints on the recipes of dust modelling and star formation. None of the \flares\, regions have galaxies that have moved into the passive regime at $z>7$, thus it is not surprising that the double power-law performs better at the higher redshifts. 

\subsubsection{Comparison with Observations and Models}\label{sec: PhotProp.UVLF.Compare}
\begin{figure*}
	\centering
	\includegraphics[width=\textwidth]{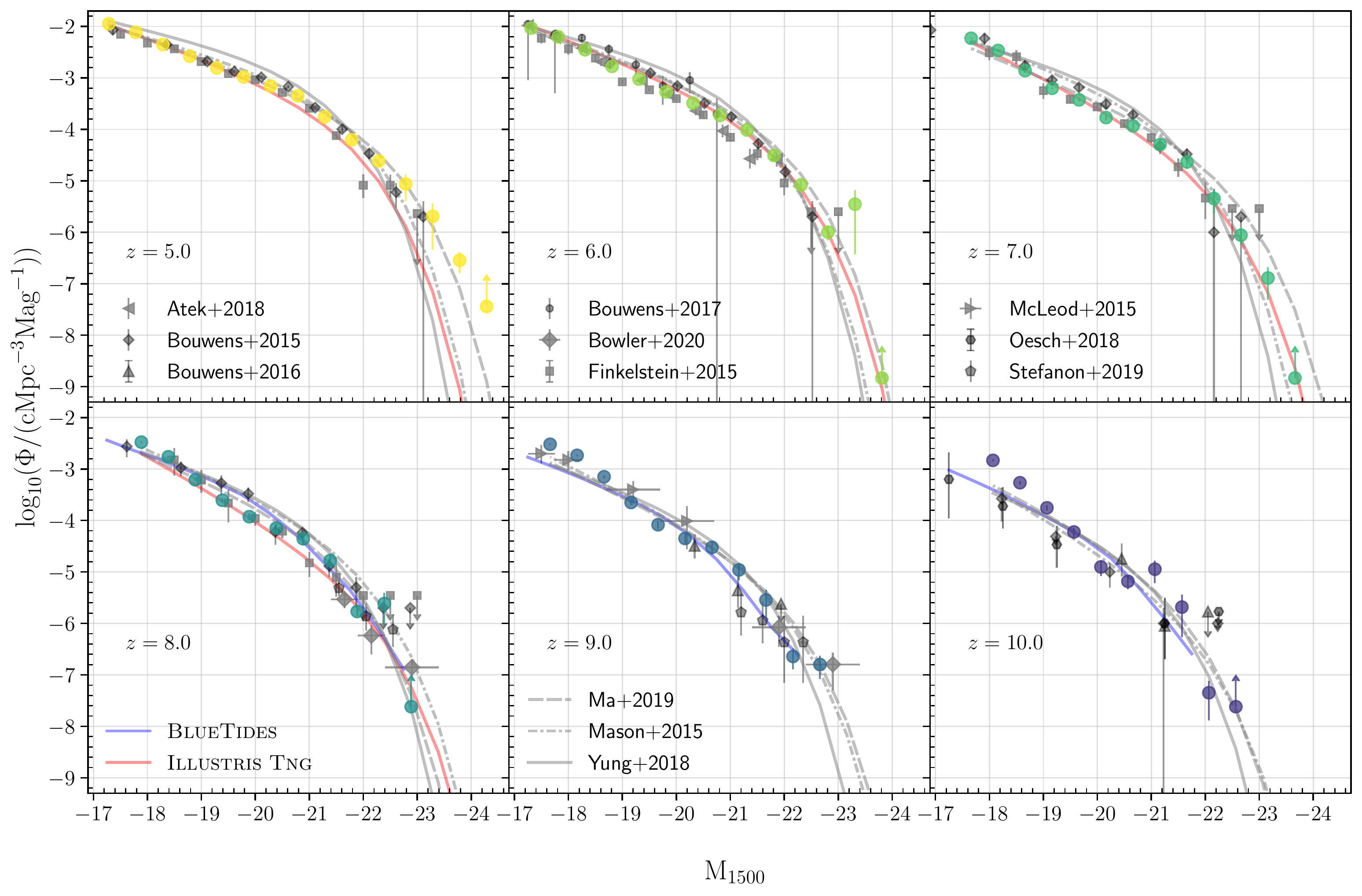}
	\caption{UV LF of the \flares\, galaxies, represented by the large coloured dots for z $\in$ [5,10]. Error bars denote the Poisson $1\sigma$ uncertainties for each bin from the simulated number counts for the dust attenuated UV LF. Observational data from \protect\cite{Bouwens_2015a,McLeod2015,Finkelstein2015,Bouwens_2016,Bouwens2017,Oesch_2018,Atek2018,Stefanon2019,Bowler2020} are plotted as well as the binned luminosities from \bluetides\,\protect\citep{Wilkins2017} and the Schechter fits from \protect\cite{Mason2015,Ma2019, Yung2019a}; \textsc{Illustris Tng} \protect\citep[Model-C from][]{Vogelsberger2020} are shown for comparison.  \label{fig: UVLF Compare}} 
\end{figure*}
In Figure~\ref{fig: UVLF Compare} the UV LF of \flares\, galaxies is compared to observational values from
\cite{Bouwens_2015a,McLeod2015,Finkelstein2015,Bouwens_2016,Bouwens2017,Oesch_2018,Atek2018,Stefanon2019,Bowler2020}. The Schechter as well as the double power-law fit to the \flares\, population is also shown. 

The UV LF relation of the \flares\, galaxies at all redshift is in good agreement within the observational uncertainties. It should also be noted that the uncertainties in the observations gets progressively larger with increasing redshift and some of the number densities at the bright end are upper limits. We slightly over-predict the number density of galaxies at $z=10$ at the faint-end. However, the observations at $z=10$ are limited by the Hubble Space Telescope's capability to detect galaxies, and hence the \cite{Oesch_2018} study contain a total of only 9 galaxies. This will change with the imminent launch of \jwst\,, which will be able detect a larger sample of galaxies and bring tighter constraints.

In Figure~\ref{fig: UVLF Compare}, we also plot the binned luminosities from \bluetides\,\citep{Wilkins2017} and the Schechter function fits from \cite{Mason2015}, \textsc{Fire-2} \citep{Ma2019}; \textsc{SantaCruz} SAM \citep{Yung2019a}; \textsc{Illustris-Tng} \cite[Model-C from][]{Vogelsberger2020}. As can be seen the fit is similar to others from literature, and only starts to diverge slightly at $z\ge8$, with \flares\, having a lower number density at the bright end compared to the Schechter fits from \cite{Mason2015,Ma2019}. Modelling differences across the studies or the larger dynamic range probed by \flares\, is a possible explanation for this deviation. With respect to \bluetides\,, a comparison of data have shown us that the most massive galaxies in \flares\, are more metal rich by $\sim$0.1 dex. This results in increased dust attenuation in \flares\, compared to \bluetides\, in , and thus cause differences in the observed UV continuum, attenuation and line luminosity values presented in the next sections. However, a direct comparison to \cite{Wilkins2017,Wilkins2020}, which also implemented a similar line-of-sight attenuation model, is not possible due to the difference in the modelling approach, namely the implementation of birth cloud attenuation and the dependence on an evolving DTM ratio. 

In Figure~\ref{fig: fit param evo} we also plot fit parameters from other studies of simulations \citep{Mason2015,Wilkins2017,Yung2019a,Vogelsberger2020} as well as observations \citep{Finkelstein2015,Bowler2015,Bowler2020}.  There exists degeneracies between the fit parameters \citep[see][]{Robertson2010a}, and these depend upon the dynamic range and the statistics of the galaxy population. \flares\, probes higher density regions, and can therefore better sample the bright end as well as the knee of the function. Thus it is not straightforward to compare fit parameters from different studies. 

\subsection{UV continuum slope ($\beta$)}\label{sec: PhotProp.beta}
\begin{figure*}
	\centering
	\includegraphics[width=0.97\textwidth]{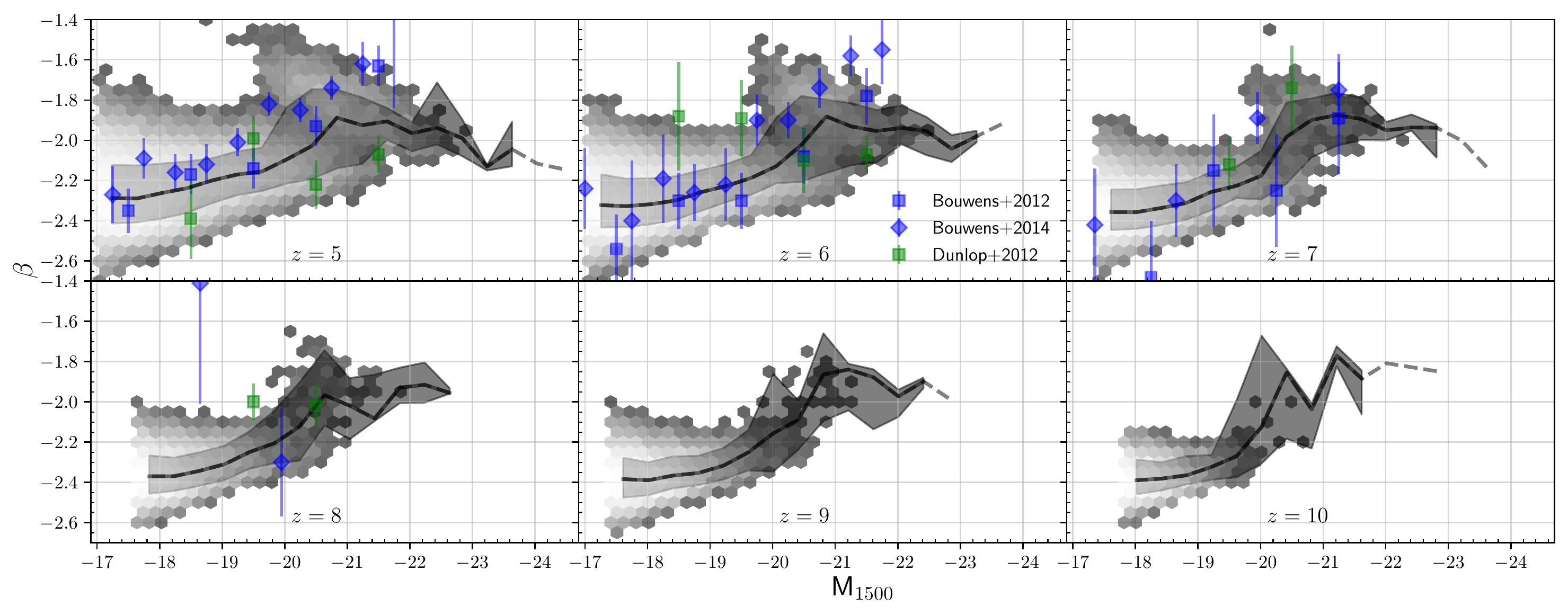}
	\caption{UV continuum slope, $\beta$, against the UV magnitude for $z\in[5,10]$. The solid dashed line is the weighted median of the sample, with the shaded region indicating the weighted  84$^{\text{th}}$ and 16$^{\text{th}}$ percentiles. The hexbin denotes the distribution of our sample. We only plot bins with more than 5 data points. Plotted alongside are observational values from \protect\cite{Dunlop2012,Bouwens2012b,Bouwens2014a}.\label{fig: beta}} 
\end{figure*} 
\begin{figure*}
	\centering
	\includegraphics[width=0.95\textwidth]{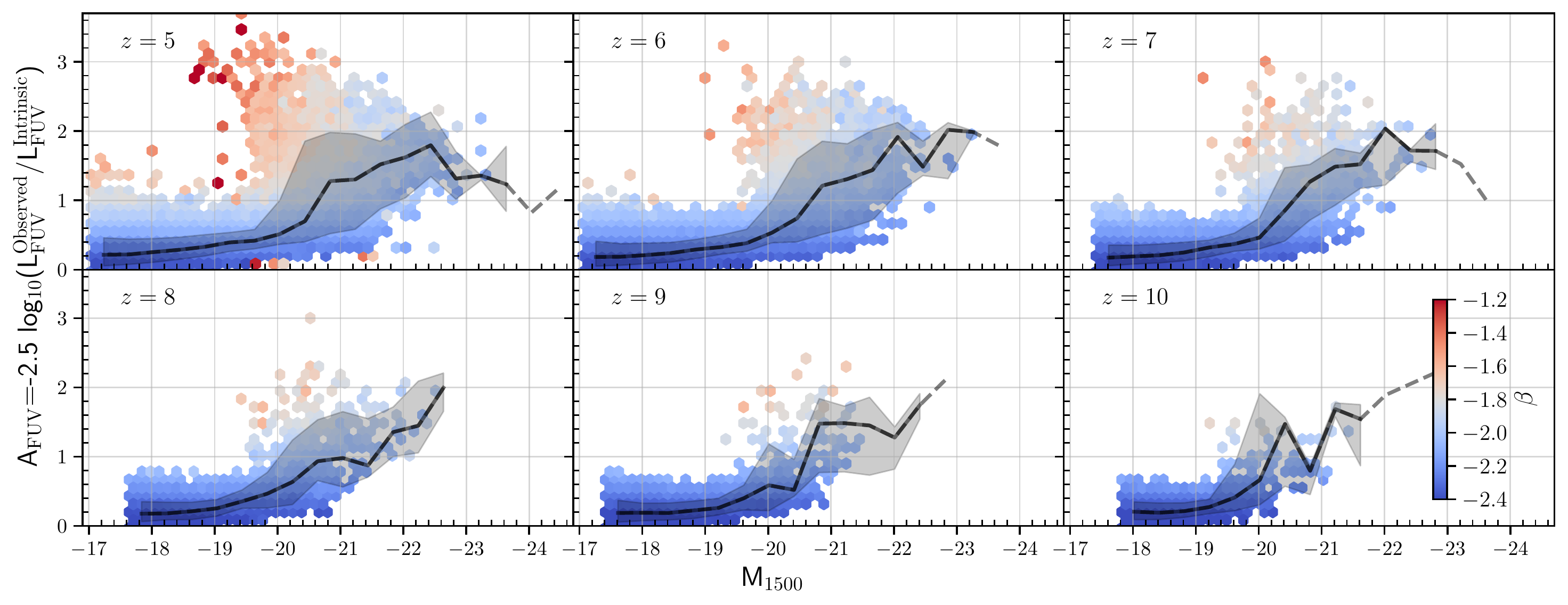}
	\caption{The attenuation in the FUV against the observed UV magnitude for $z\in[5,10]$. The solid and dashed black line is the weighted median of the sample, with the shaded region indicating the weighted 84$^{\text{th}}$ and 16$^{\text{th}}$ percentiles. The dashed line is for bins that have less than 3 data points. The hexbin denotes the distribution of our sample, coloured by the median $\beta$ value in the hexbin. \label{fig: att_lfuv}} 
\end{figure*} 
\begin{figure*}
	\centering
	\includegraphics[width=0.95\textwidth]{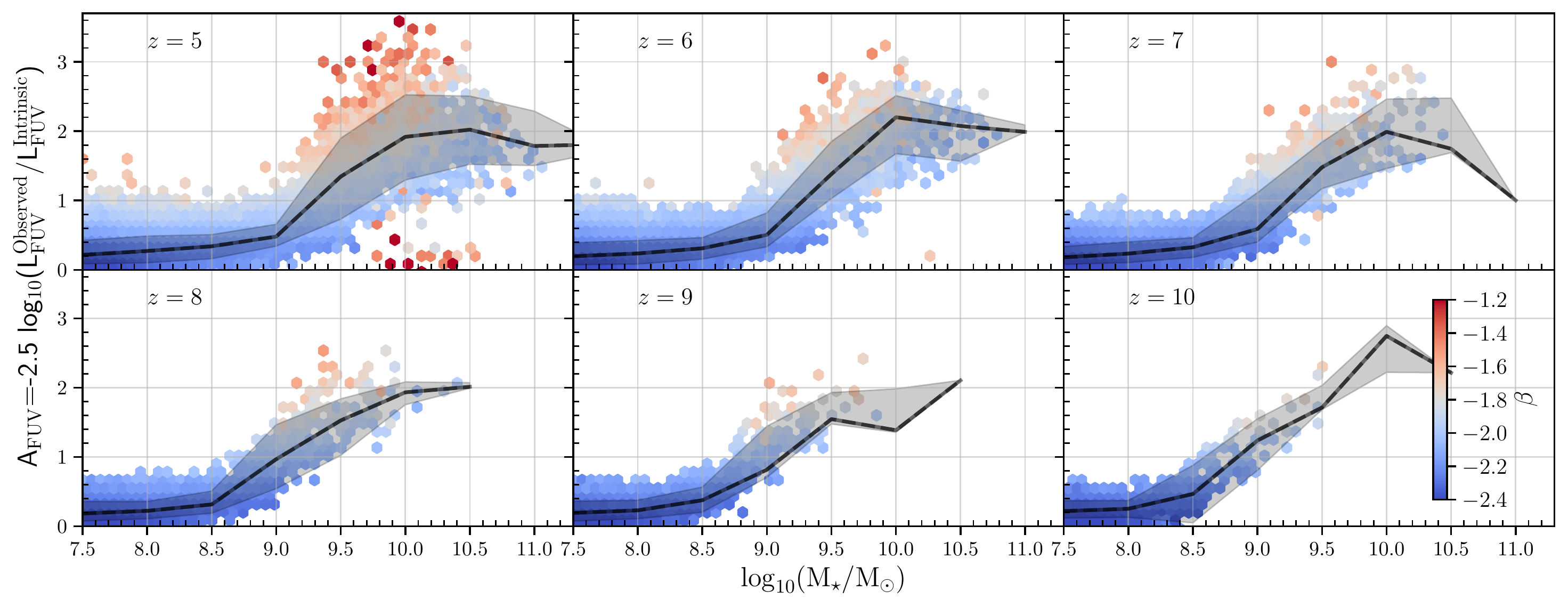}
	\caption{The attenuation in the FUV against the galaxy stellar mass for $z\in[5,10]$. The solid and dashed black line is the weighted median of the sample, with the shaded region indicating the weighted 84$^{\text{th}}$ and 16$^{\text{th}}$ percentiles. The hexbin denotes the distribution of our sample, coloured by the median $\beta$ value in the hexbin. \label{fig: att_Mstar}} 
\end{figure*} 
The UV continuum slope $\beta$, defined such that f$_{\lambda}\,\propto\,\lambda^{\beta}$ \citep{Calzetti1994}, is commonly used as a tracer of the stellar continuum attenuation. At high redshifts, the rest-frame UV becomes accessible to optical/near-IR instruments. This has been studied by different groups \cite[\eg][]{Stanway2005,Wilkins2011b,Dunlop2012,Finkelstein2012,Bouwens2014a,Bhatawdekar2020} as it is accessible due to deep near-IR observations using the Wide Field Camera 3 (WFC3) on the \textit{Hubble Space Telescope}. These studies have shown that $\beta$ is particularly sensitive to the metallicity, age, and especially the dust content within a galaxy, and thus it is a useful quantity to check the reliability of theoretical models. However, it is important to note that $\beta$ is also strongly dependent upon the modelling assumptions like the choice of the IMF, SPS model, dust modelling and extinction law.   

Figure~\ref{fig: beta} plots the value of $\beta$ against the UV luminosity of the galaxies in \flares. Observational values of $\beta$ from \cite{Dunlop2012,Bouwens2012b,Bouwens2014a} are plotted alongside for comparison. It should be noted that the observational data shows a lot of scatter and the different datasets do not show the same trends. Our weighted median of $\beta$'s  match observational values for almost all luminosities. At the bright end, M$_{1500}<-20$ the \cite{Bouwens2012b,Bouwens2014a} data predict much steeper $\beta$'s compared to our results, which start to flatten while \cite{Dunlop2012} shows lower values. This could be due to the choice of our extinction curve, a steeper/shallower curve will make for a steeper/shallower relation. The $\beta$ values are an excellent constraint on the theoretical extinction curves, giving insights into the dust properties within the galaxy \cite[see][]{Wilkins2012,Wilkins2013,Salim2020}. We examine a few extinction curves from the literature (namely the Calzetti \citep{Calzetti2000}, Small Magellanic Cloud \citep{Pei1992} and the curve used in \citealt{Narayanan2018}) in Appendix~\ref{app:extinction_curves} and plot the effect it has on the  UV continuum relation in Figure~\ref{fig: ext curves beta att} (left panel). We find that the \flares\, galaxies prefer a steeper extinction curve similar to the SMC in order to reproduce UV continuum observations. It is interesting to note in this context that \cite{Ma2019} probed the IRX-$\beta$ relation in the \textsc{FIRE-2} simulation suite using the radiative transfer code \textsc{Skirt} \citep{Baes2015}, and obtained a relation which is broadly in agreement with using a simple screen model with the SMC extinction curve.

\cite{Shen2020} showed the relation between M$_{\mathrm{UV}}-\beta$ (their Figure 9), obtained from applying \textsc{Skirt} on the \textsc{Illustris-Tng} suite of simulations. Similar to what is seen in Figure~\ref{fig: beta}, the $\beta$ values start to flatten at the bright end. \cite{Wu2020}, using the \textsc{Simba} simulation suite, capture a similar relation, albeit with a higher normalisation using the \cite{Calzetti2000} extinction law. \textsc{Simba} implements a self-consistent dust model, which allows them to infer the dust column density directly and use this in their line-of-sight dust attenuation model. They find that dust attenuation becomes important at M$_{1500}<-18$, while in \flares\, it starts only at M$_{1500}\lesssim-21$ at $z=6$. This extra dust extinction could explain the difference in normalisation seen.

In Figure~\ref{fig: att_lfuv} we plot the attenuation in the UV against the UV luminosity, in hexbins coloured by the median $\beta$ value. The value of the attenuation provides insight into the amount of obscured star formation that is going on in galaxies (also see \S\ref{sec: sfr}). Overall, brighter galaxies suffer more attenuation, which is expected as they have had more time to produce stars thus enriching the ISM. We can also see that there is a sudden increase in the UV-attenuation for galaxies brighter than -20 magnitude, pointing towards the rapid build-up of dusty galaxies in this regime. The figure also shows that many of the galaxies at the bright end are not the most attenuated ones. These are the galaxies that have enjoyed a recent burst of star formation and have not had time to enrich the ISM with dust. Another alternative is stellar migration \citep[see][]{furlong_evolution_2015}, with some stars moving radially outwards, thus subject to reduced dust attenuation depending on the viewing angle or geometry. Some recent ALMA studies at high redshift \cite[\eg][]{Bowler2018} have also found galaxies having a heavily dust-obscured and an unobscured component. The variation of dust attenuation within a galaxy as well as the viewing angle will be explored in a future work. \cite{Ma2020}, using the FIRE-2 simulation, studied the escape fraction of ionizing photons across different resolutions, and found that the lowest resolution run had a lower escape fraction compared to the higher resolutions. In a future study we plan to explore the effect of dust attenuation with resolution on our dust model. In Figure~\ref{fig: att_lfuv_intrinsic} we plot the attenuation as a function of intrinsic FUV luminosity. This provide more insights into the features seen in Figure~\ref{fig: att_lfuv}; in general, intrinsically brighter galaxies are more attenuated. A comparison also reveals that many of the intrinsically bright galaxies, since they are dusty, are not the brightest galaxies observed in the UV. The relations presented above are also in agreement with the A$_{\mathrm{UV}}-\mathrm{M}_{\star}$ and the A$_{\mathrm{UV}}-\beta$ relations presented in \cite{Shen2020} (their Figures 10 and 11) at $z\le6$.

We also plot the attenuation as a function of galaxy stellar mass in Figure~\ref{fig: att_Mstar}. Features similar to the plots described earlier are seen here as well, with a flattening of the relation at the low mass end ($\lesssim10^{8.5}$M$_{\odot}$), and rapid steepening afterwards. As seen in local observations our values do not exhibit a large scatter at the low mass end. This scatter at low redshift can be explained by varying dust content and star-dust geometries of the galaxies. High resolution simulations such as FIRE-2 \citep[see][]{Ma2019} also see a flattening of the FUV attenuation at the low mass end, with more scatter, possibly due to the low number galaxies produced at the massive end. 

We have examined the few galaxies at $z=5$ that have very low attenuation, but have high $\beta$ values (also seen in Figure~\ref{fig: att_Mstar}). They also are intrinsically very bright (see Figure~\ref{fig: att_lfuv_intrinsic}). These are galaxies that are identified to be in the passive regime, whose specific star formation rate was calculated to be $\lesssim$ 1/(3$\times$H(z)), where H(z) is the Hubble constant at $z=5$. We will be studying this population in more detail in a future work (Lovell et al. in prep). 

\subsection{Effect of environment}\label{sec: PhotProp.Env}

\begin{figure*}
	\centering
	\includegraphics[width=0.97\textwidth]{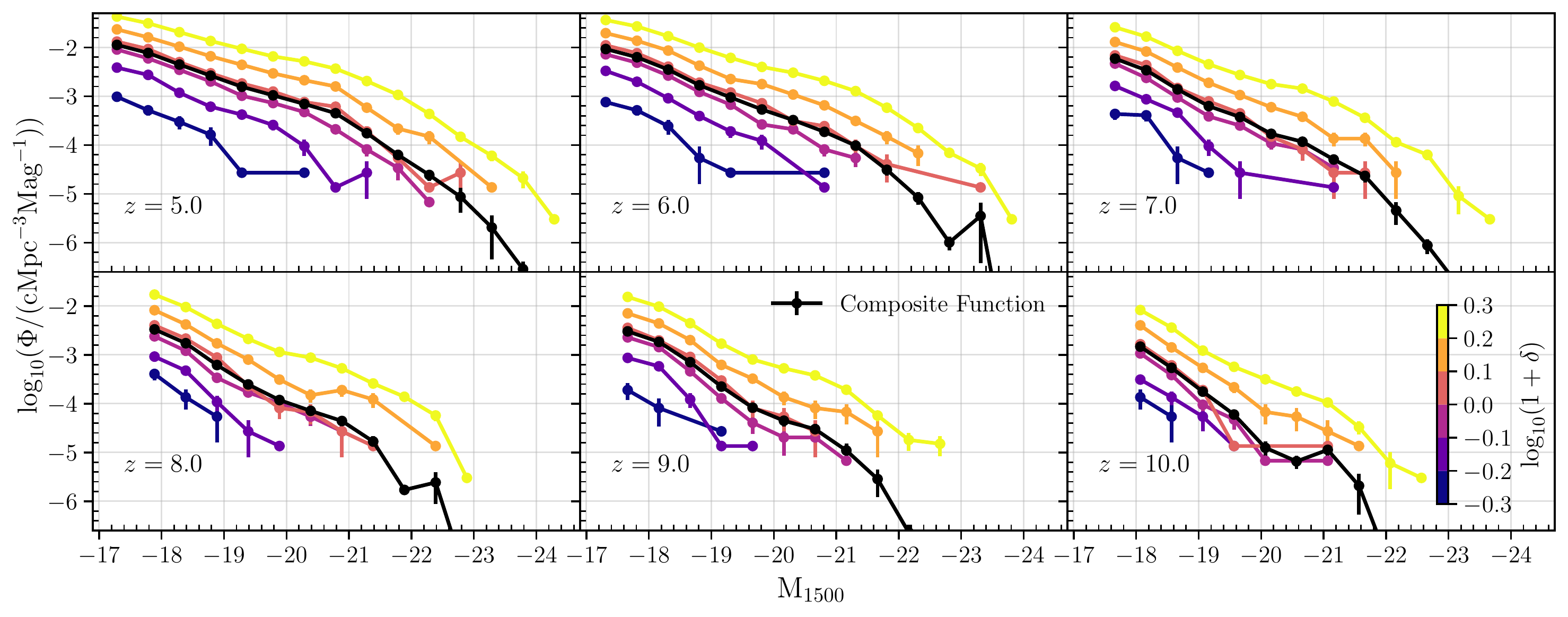}
	\caption{The \flares\, UV LF for $z \in [5,10]$ split by binned log-overdensity. Error bars denote the Poisson $1\sigma$ uncertainties for each bin from the simulated number counts. \label{fig: UVLF env}} 
\end{figure*}
The \flares\, probes galaxies that reside in a wide range of environments allowing us to analyse the effect environment has on their observed properties. In Figure~\ref{fig: UVLF env} we look at how the UV LF varies as a function of overdensity for $z \in [5,10]$. Here we have plotted the UV LF in 6 bins of log$_{10}$($1+\delta$), where $\delta$ is the overdensity. As expected the number density of galaxies increases with increasing overdensity and the brighter galaxies reside predominantly in denser environments. Similar behaviour has been seen in measurements of the UV LF in high-redshift galaxy protoclusters \citep{ito_uv_2020}. The normalisation shows a variation of $\sim2$ dex from the lowest to the highest density environment probed in \flares, much greater than the 0.5 dex variation in density itself. The composite distribution function closely follows that of mild overdensity, $\log_{10}(1+\delta)\in 0 - 0.1$, with the contribution to the bright end coming only from the densest environments. 

As can be seen from Figure~\ref{fig: UVLF env} the shape of the luminosity function is similar across various environments with no significant variation in the knee of the function. There is a hint of a double Schechter shape, being strongest in intermediate to lower density environments at high redshift. This could be due to the different assembly histories of galaxies driven by the environment. The effect of environment on assembly history as well as on astronomical surveys will be probed in a future work (Thomas et al. in prep). 

We have also looked at the UV continuum slopes as well as the attenuation in the far-UV as a function of environment similar to the method described above. We find no dependence on overdensity for these galaxy properties.

\subsection{Line Luminosities and Equivalent Widths}\label{sec:PhotProp.lines}

\begin{figure*}
	\centering
	\includegraphics[width=\textwidth]{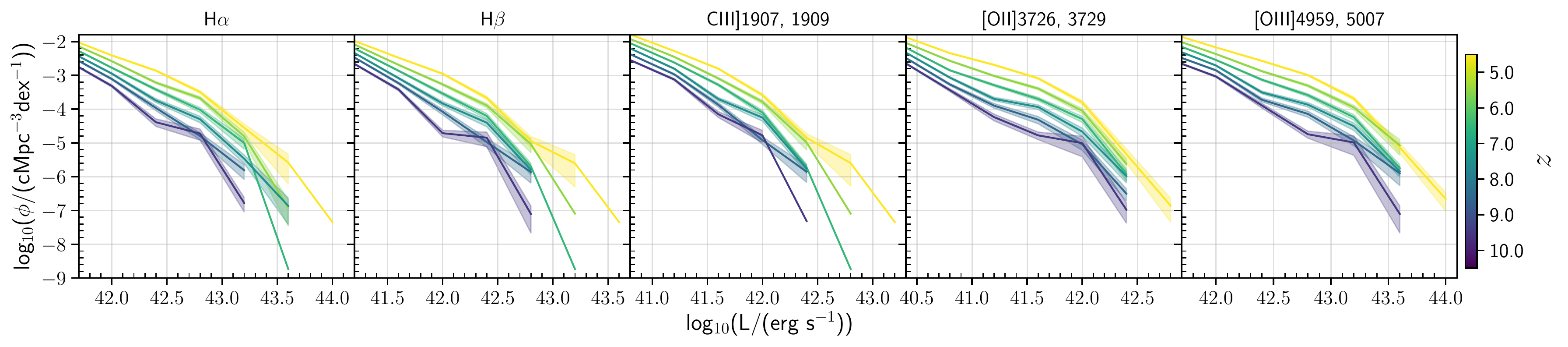}
	\hspace*{-1.1cm}\includegraphics[width=0.94\textwidth]{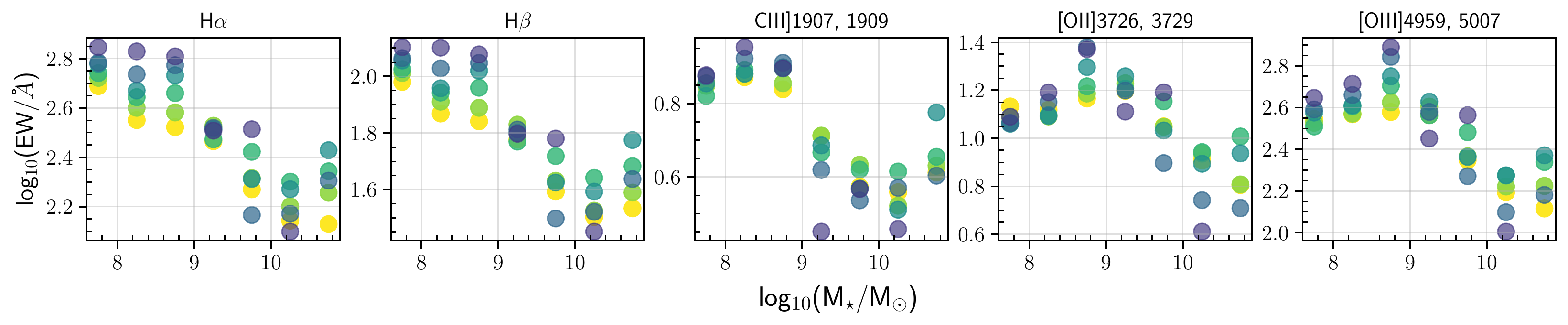}
	\hspace*{-1.1cm}\includegraphics[width=0.94\textwidth]{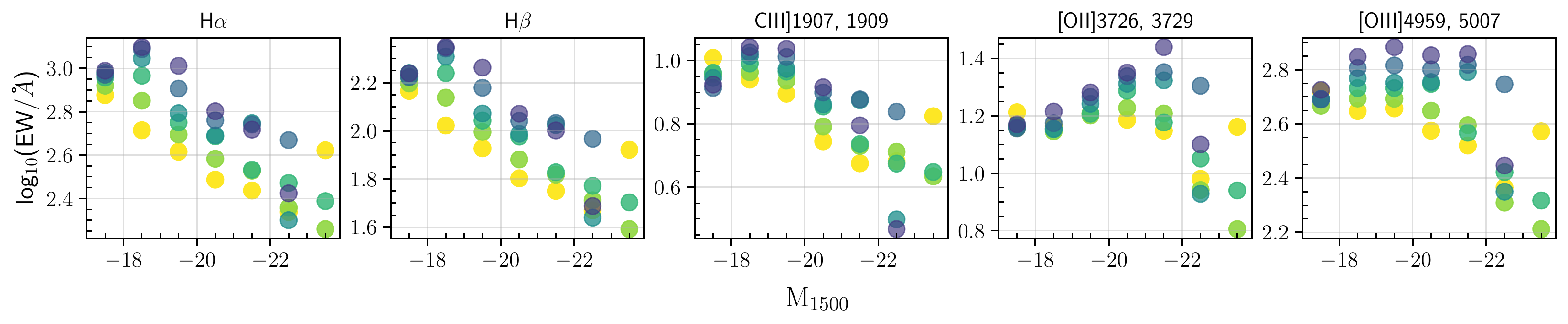}
	\caption{Predictions for the properties of 6 prominent UV and optical lines in \flares\, for $z\in[5,10]$. The colour bars for the different redshifts are shown in the rightmost panel. In the top panel we show the dust-attenuated luminosity functions for each line, with the shaded region representing the 1$\sigma$ Poisson uncertainties. Middle panel shows the evolution of the weighted median equivalent widths of these lines in stellar mass bins. Bottom panel shows the weighted median equivalent widths as a function of FUV luminosity. \label{fig: LF_line_evo}} 
\end{figure*} 
\begin{figure*}
	\centering
	\includegraphics[width=0.97\textwidth]{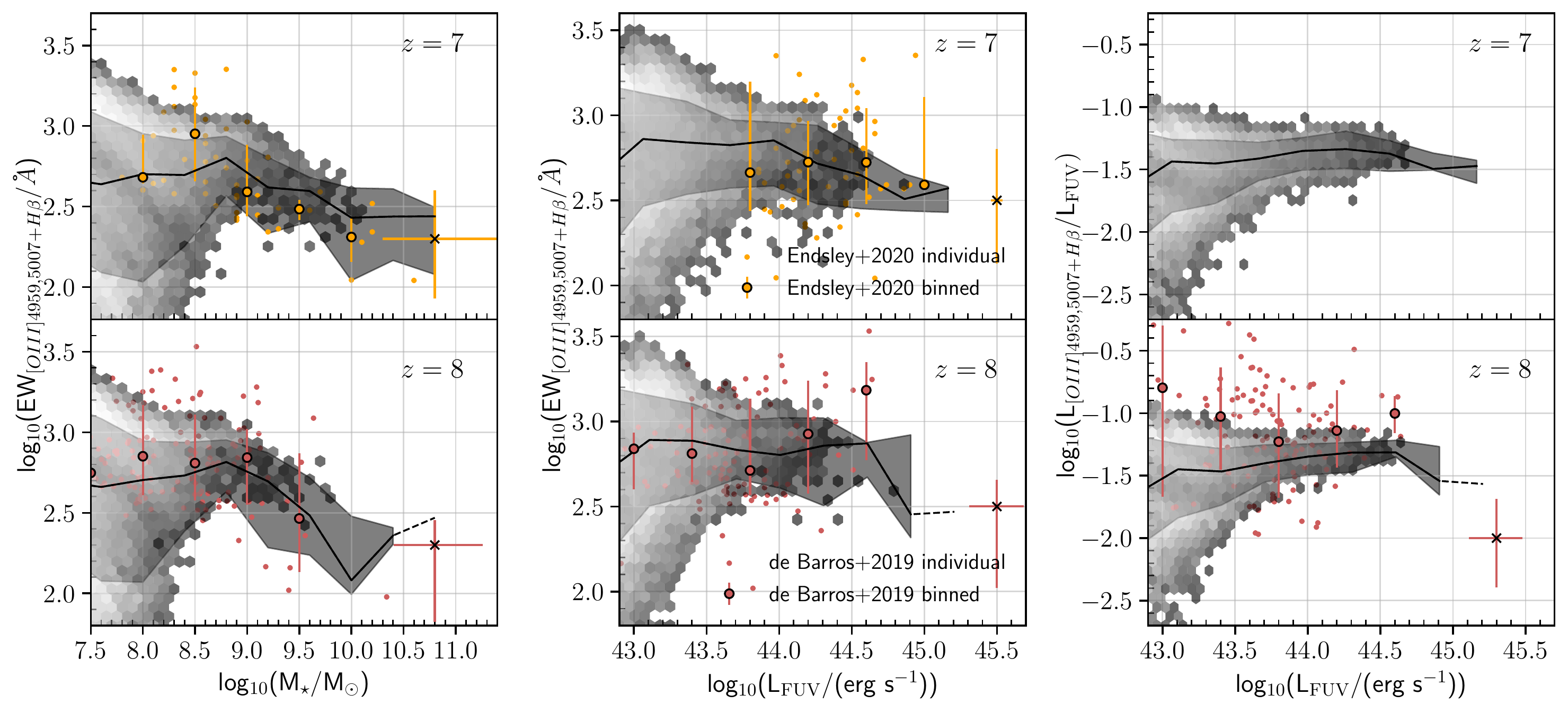}
	\caption{Left: Predicted distribution of combined H${\beta}$ and [OIII]$\lambda$4959,5007 equivalent widths and stellar masses for \flares\, galaxies at $z\sim7,8$. Middle: Predicted distribution of combined H${\beta}$ and [OIII]$\lambda$4959,5007 equivalent widths to the far-UV luminosity of \flares\, galaxies at $z\sim7,8$. Right: Predicted distribution of the H${\beta}$ and [OIII]$\lambda$4959,5007 line luminosities to the far-UV luminosity and far-UV luminosities of \flares\, galaxies at $z\sim7,8$. The solid line is the weighted median of the sample, with the shaded region indicating the weighted 84$^{\text{th}}$ and 16$^{\text{th}}$ percentiles. The hexbin denotes the distribution of our sample, only plotted are bins with more than 5 data points. The small circles show the individual measurements from \protect\cite{deBarros19_OIIIHbeta,Endsley2020} while the large points denote the median value in bins of stellar mass and far-UV luminosities respectively. The errorbars centered on the cross shown at the bottom-right gives the median errors on the observational data.\label{fig: OIII_z7_8}} 
\end{figure*} 
\begin{figure}
	\centering
	\includegraphics[width=0.95\columnwidth]{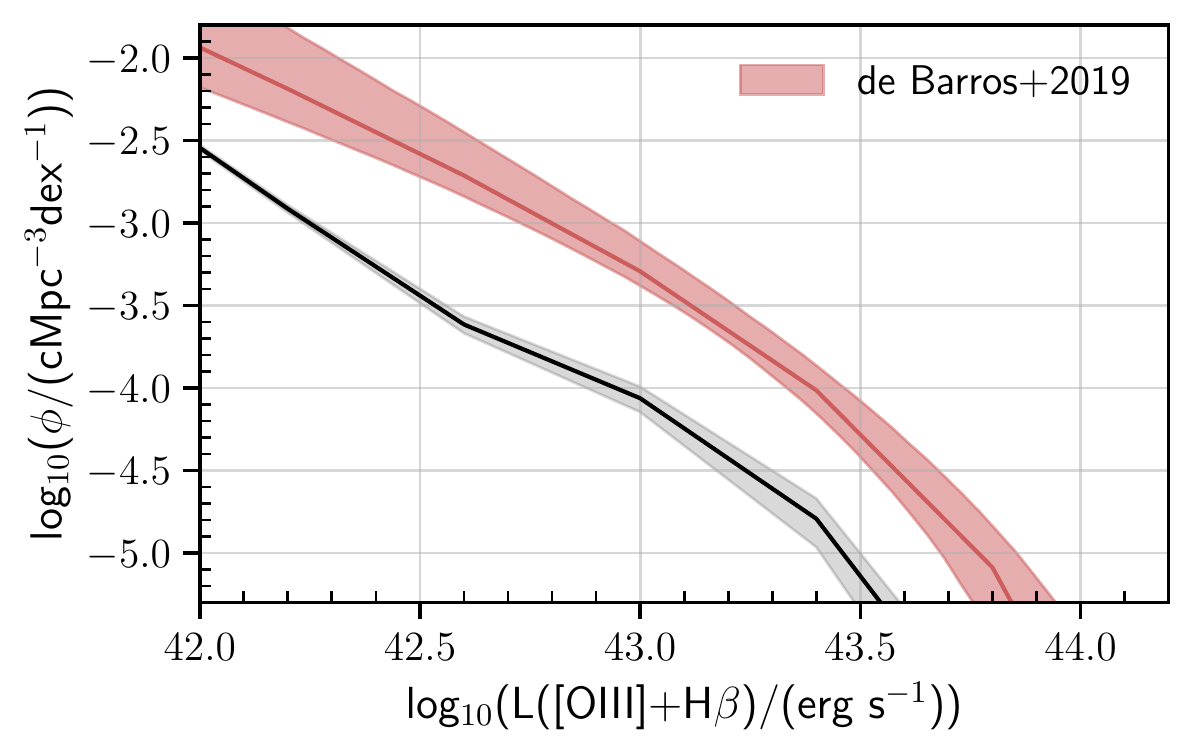}
	\caption{The \protect\cite{deBarros19_OIIIHbeta} and predicted combined H$\beta$ and [OIII]$\lambda$4959,5007 line luminosity function of \flares\, galaxies at $z\sim8$. \label{fig: OIII_LF_8}} 
\end{figure}
\begin{figure}
	\centering
	\includegraphics[width=0.97\columnwidth]{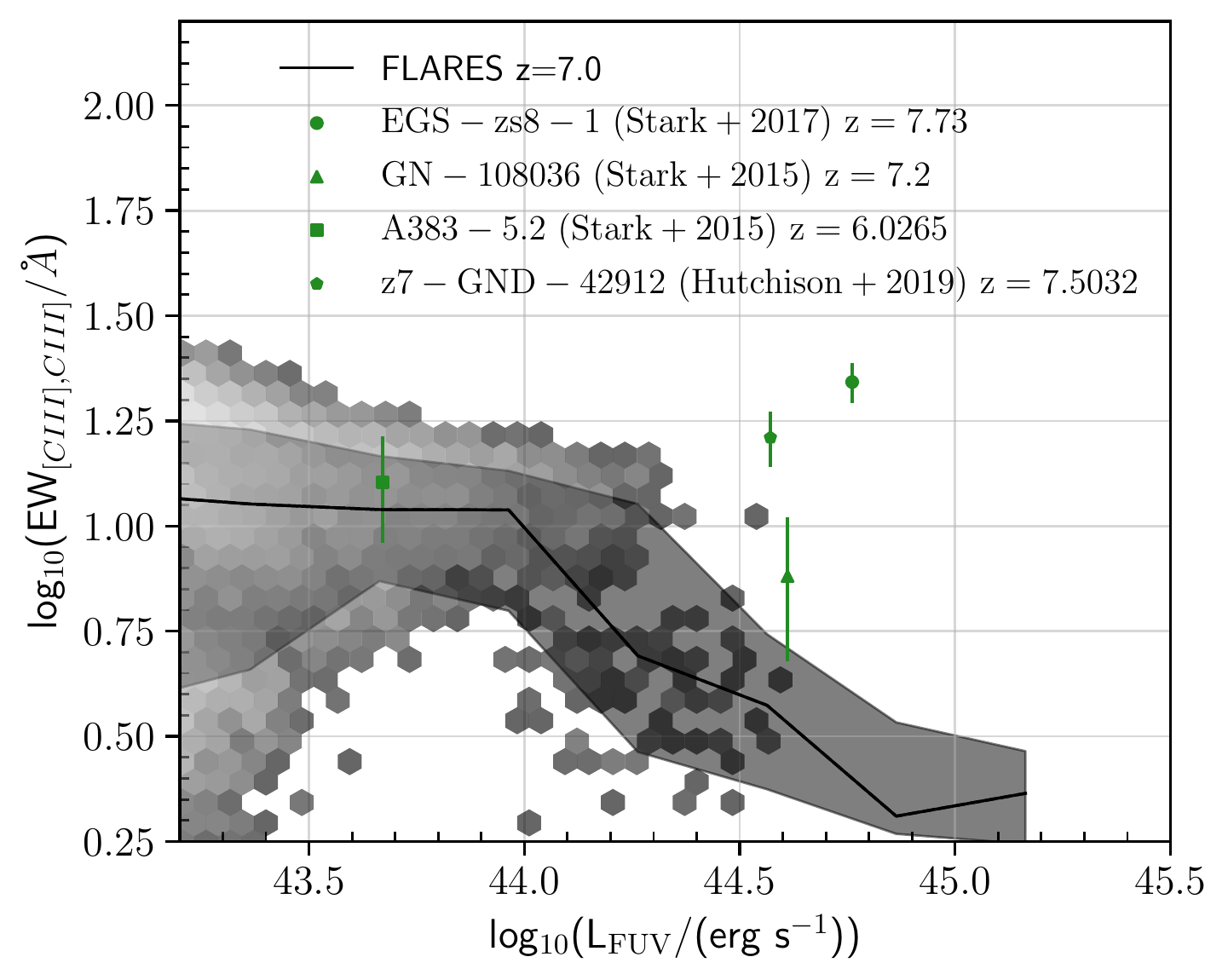}
	\caption{Predicted [CIII]$\lambda$1907,$\lambda$1909 line equivalent widths of \flares\, galaxies at $z\sim7$. The solid line is the weighted median of the sample, with the shaded region indicating the weighted 84$^{\text{th}}$ and 16$^{\text{th}}$ percentiles. The hexbin denotes the distribution of our sample, only plotted are bins with more than 5 data points. Plotted alongside are observational values from \protect\cite{Stark2015,Stark2017,Hutchison2019}. \label{fig: CIII_EW}} 
\end{figure}
In this section we will present some of the nebular emission line properties and compare them to some of the available observational constraints. 
 
We present predictions for 6 prominent nebular lines or doublets in the UV in Figure~\ref{fig: LF_line_evo}. The top panel shows the evolution of the line luminosity function with redshift, for $z\in[5,10]$. The overall shape of the function is similar to the UV luminosity function of galaxies and can be approximated by a Schechter function at these redshifts. The LF of the lines evolves with redshift, with almost 3 dex in value near the knee of the function. We also present predictions for the evolution of the weighted median equivalent widths of these lines as a function of stellar mass (middle panel) and far-UV luminosity (bottom panel) with redshift in Figure~\ref{fig: LF_line_evo}. For galaxies with similar stellar mass the equivalent width mostly increases with increasing redshift, indicating that they have harder ionising photons from their younger stellar population with more massive stars. There is also the effect of metallicity on the line width, causing them to drop quickly at higher stellar masses in case of the hydrogen recombination lines, while the other lines peak around 10$^9$M$_{\odot}$ and then fall rapidly. In case of the far-UV, the relationship with metallicity is not correlated in the same way as stellar mass and hence interpretation is harder. But in most cases this also shows increasing equivalent widths at higher redshifts for fixed far-UV luminosity. This behaviour is in agreement with that seen from the \bluetides\, simulation presented in \cite{Wilkins2020}.

Both \cite{deBarros19_OIIIHbeta,Endsley2020} have combined broadband photometry from \textit{Hubble} and \textit{Spitzer} observations to constrain the prominent H$\beta$ and [OIII]$\lambda$4959,5007 lines at $z\sim 7,8$. In Figure~\ref{fig: OIII_z7_8} we plot the combined values of [OIII]$\lambda$4959,5007 and H$\beta$ line luminosities as well as the equivalent widths (EWs) of \flares\, galaxies at $z=7,8$ against these observational data sets. As can be seen from the figure, in the case of the equivalent width measurements plotted against the stellar mass (left panel) or FUV luminosity (middle panel), the weighted median closely follows the observations. However, it should be noted that our modelling does fail to reproduce some of the larger values of the EW measurements. In case of the line luminosity normalised by the far-UV luminosity (right panel), \flares\, lies $\sim 0.3$ dex below the observational data from \cite{deBarros19_OIIIHbeta}. We also compare the [OIII]$\lambda$4959,5007 luminosity function as predicted by \cite{deBarros19_OIIIHbeta} at $z=8$ to \flares\, in Figure~\ref{fig: OIII_LF_8}. Our result is offset by $\approx0.6$ to lower number densities or by $\approx0.4$ to lower luminosities. The cause of this offset could be due to the relation used by \cite{deBarros19_OIIIHbeta} to convert the observed far-UV LF to a line luminosity LF. A similar feature is also seen in the $z=8$ [OIII]$\lambda$4959,5007 luminosity function from the \textsc{Illustris-Tng} simulation  presented in \cite{Shen2020} (their Figure 5), with marginal consistency at the bright end ($>43.5$ erg/s). \cite{Wilkins2020} also show an underprediction of the luminosity function at $z=8$.

We also show the predicted [CIII]$\lambda$1907,$\lambda$1909 line equivalent widths of \flares\, galaxies at $z\sim7$ against observations from \cite{Stark2015,Stark2017,Hutchison2019} in the redshift range of $6-8$ in Figure~\ref{fig: CIII_EW}. A similar feature is seen here as well where we underpredict some high-EW measurements at the most luminous end. An explanation of this discrepancy could be due to the assumptions in the nebular emission modelling like the nebular hydrogen density or ionisation parameter \citep[see Section 3.4 in][for more details]{Wilkins2020} as well as contributions from AGN which we have not considered in this work. Future direct emission line measurements from \jwst\, and other facilities will help to constrain this observational space and thus better understand this discrepancy. 

\section{SFR distribution functions}\label{sec: sfr}

\begin{figure*}
	\centering
	\includegraphics[width=0.97\textwidth]{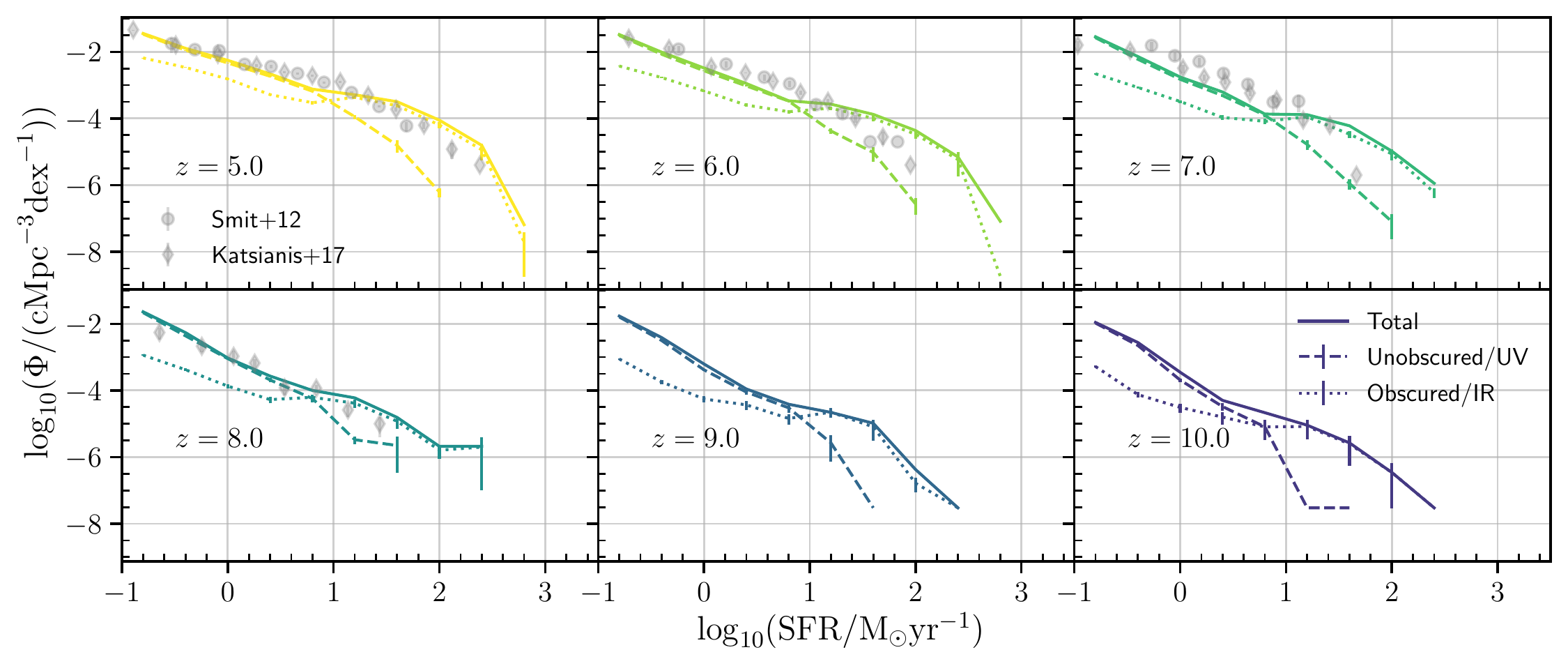}
	\caption{\flares\, composite galaxy total (solid), obscured (dotted) and unobscured (dashed) star formation rate function for z $\in$ [5,10]. The 1-$\sigma$ Poisson uncertainties for the obscured and unobscured star formation rate function are also plotted. For comparison the dust-corrected SFRF from \protect\cite{smit_star_2012,katsianis_evolution_2017} is also shown. \label{fig: sfrf}} 
\end{figure*}
\begin{figure}
	\centering
	\includegraphics[width=0.97\columnwidth]{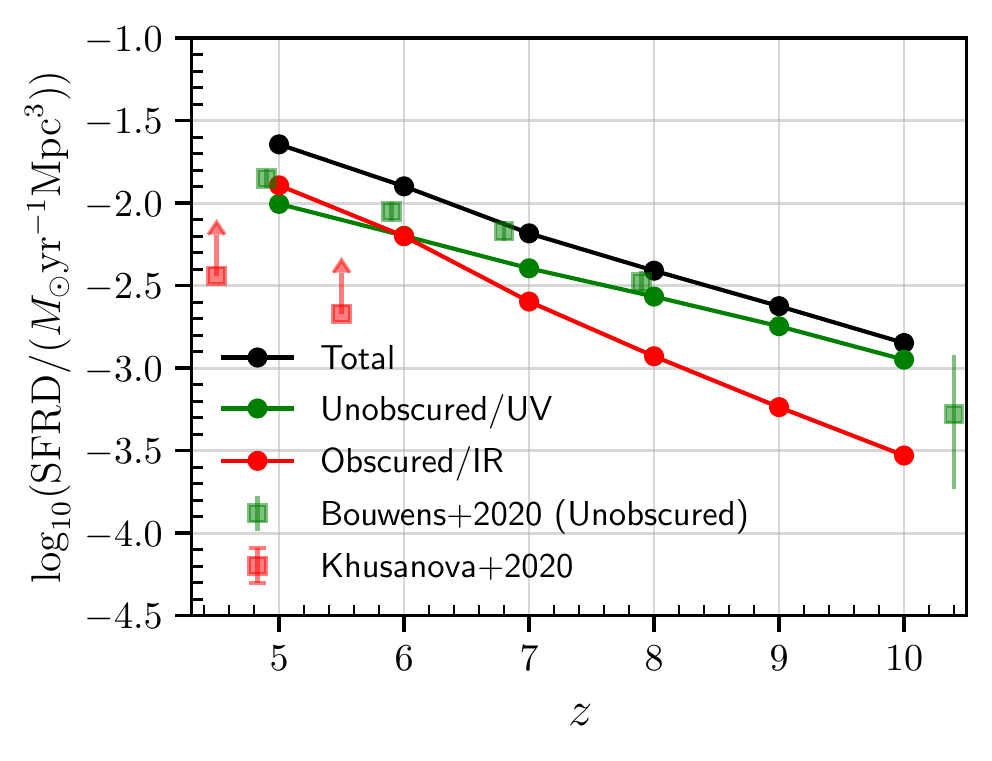}
	\caption{\flares\, composite galaxy total (solid), obscured (dotted) and unobscured (dashed) star formation rate density for z $\in$ [5,10]. For comparison the uncorrected SFRD or SFRD$_{\textrm{UV}}$from \protect\cite{Bouwens2020} (obtained from UV luminosity scaling relations) and SFRD$_{\textrm{IR}}$ from \protect\cite{Khusanova2020} (which are lower limits) is also shown. \label{fig: sfrd}} 
\end{figure}
The instantaneous SFR distribution function of the \flares\, galaxies was already presented in \citetalias{lovell2020}, which followed a double Schechter form and provided a good match to the observed values. In this section we look at the relative contribution of the obscured and unobscured/uncorrected star formation rate in \flares\,. We compute the fraction of obscured star formation or infrared star formation rate, f$_{\mathrm{obsc}}$ going on in any given galaxy by using the attenuation in the far-UV, A$_{\mathrm{FUV}}$. It is computed as 
\begin{gather}\label{eq: f_obsc}
	f_{\mathrm{obsc}} = 1-\frac{L_{\mathrm{FUV}}^{\mathrm{Observed}}}{L_{\mathrm{FUV}}^{\mathrm{Intrinsic}}} = 1 - 10^{-A_{\mathrm{FUV}}/2.5}\:,
\end{gather}
with f$_{\mathrm{unobsc}}=1-$f$_{\mathrm{obsc}}$ the fraction of unobscured star formation rate. Using this prescription, the rate of obscured (infrared) and unobscured (far-UV) star formation rate are f$_{\mathrm{obsc}}\times$SFR and f$_{\mathrm{unobsc}}\times$SFR, respectively. This would differ slightly from the observed calibration, where the obscured and unobscured SFRs are obtained by combining the total IR and observed UV luminosities with a theoretically motivated calibration \citep[\eg][]{kennicutt_jr_star_2012}. We use the SFR of a galaxy averaged over the star particles that were formed in the last 100 Myr. These would closely resemble SFRs inferred observationally from the UV/IR, rather than ones that were obtained by emission line calibrations.  

Figure~\ref{fig: sfrf} shows the total, obscured and unobscured SFR distribution function for the \flares\, galaxies in $z\in[5,10]$. We also plot the dust-corrected SFR function from \cite{smit_star_2012,katsianis_evolution_2017} for comparison. The dust corrections are done using the IRX-$\beta$ relation established by \cite{Meurer1999}. This can be uncertain for highly star-forming systems and possibly underestimated \citep{katsianis_evolution_2017}. As can be seen, obscured star formation dominates the contribution to the total at SFRs $\gtrsim 10$ M$_{\odot}$/yr, indicating the rapid build up of dust in these extreme star forming galaxies. This directly reflects what is seen in the FUV attenuation that is presented in Figures~\ref{fig: att_lfuv}, \ref{fig: att_lfuv_intrinsic}, and \ref{fig: att_Mstar}, where there is a rapid increase in the attenuation when moving to the very-bright/massive end of the distribution.

We also look at the evolution of the total (black), obscured (red) and unobscured (green) star formation rate density (SFRD) in Figure~\ref{fig: sfrd} for galaxies with SFR $\ge0.1$ M$_{\odot}$/yr. Even though the bright end is dominated by obscured star formation at all redshifts, we find that the contribution to the total SFRD is mainly coming from unobscured star formation that takes place in low mass galaxies, or specifically from galaxies below the knee of the SFR function. The contribution of obscured star formation is $\sim40\%$ at $z=7$ and becomes almost equal at $z\sim6$. This is similar to the fraction of obscured star formation found in recent observational surveys with ALMA \citep[\eg][]{Khusanova2020}, where they predict the SFRD$_{\mathrm{IR}}$ to possibly cross the SFRD$_{\mathrm{FUV}}$ at $z>5$. \cite{Bouwens2020} also see a transition of the SFR density being primarily unobscured at $z>5$ and obscured at $z<5$. We plot these measurements for comparison in Figure~\ref{fig: sfrd}.    
\section{Conclusions}\label{sec:conc}
We have presented the photometric results from the \flare\, simulations, a suite of zoom simulations run using the \eagle\, \citep{schaye_eagle_2015,crain_eagle_2015} simulation model probing a wide range of overdensities in the Epoch of Reionisation ($z\ge5$). The wide range of overdensities sampled from a large periodic volume allows us to probe brighter and more massive galaxies in the EoR. Using a simple line-of-sight dust extinction model we retrieve the photometric properties of the galaxies in the simulation. Our main findings are as follows:
\begin{itemize}
	\item The \flares\, UV LF provides an excellent match to current observations of high-redshift galaxies. The UV LF exhibits a double power-law form at all redshifts with the Schechter form being comparable at $z=10$ from BIC. The number density of bright objects at the knee of the function increases by almost 2 orders of magnitude. At $z>8$ the number density of galaxies at the bright-end as predicted by \flares\, is less than that predicted from Schechter fits from some simulation studies. The normalisation of the UV LF is strongly dependent on the environment, with the shape being affected to a lesser extend.  
	\item The relationship between the UV continuum slope, $\beta$ and M$_{1500}$ of the \flares\, galaxies are in very good agreement with the observations. We find a flattening of the relation at the bright-end. The attenuation in the far-UV also shows a linear relationship with the observed as well as the intrinsic UV luminosity.
	\item We find good agreement of observed line luminosity and equivalent width relationship of the combined [OIII]$\lambda$4959,5007 and H$\beta$ lines as well as the CIII]$\lambda$1907,[CIII]$\lambda$1909 line equivalent widths. 
	\item The star formation in galaxies with a SFR $\gtrsim10$ M$_{\odot}$/yr is predominantly obscured and vice versa below that for the \flares\, galaxies in $z\in[5,10]$. Dust obscured star formation makes a significant contribution at these high redshifts reaching $\sim40\%$ at $z=7$, and starts dominating below $z\sim6$.
\end{itemize}
Future observations from \textit{Webb}, \euclid\, and the \textit{Roman Space Telescope} will provide further constrains on the photometric properties of these high redshift galaxies. Complimentary observations in the far-IR by \textit{ALMA} will also be instrumental in providing additional constraints on the nebular emission characteristics. We will also be investigating the emission features from PDRs in a future work. 

\section*{Acknowledgements}
We wish to thank the anonymous referee for detailed comments and suggestions that improved this paper. We thank the \eagle\, team for their efforts in developing the \eagle\, simulation code. We wish to thank Scott Kay and Adrian Jenkins for their invaluable help getting up and running with the Eagle resimulation code. We thank Desika Narayanan for providing the extinction curve used in \cite{Narayanan2018}. We thank Rebecca Bowler for providing the DPL fit parameters. This work used the DiRAC@Durham facility managed by the Institute for Computational Cosmology on behalf of the STFC DiRAC HPC Facility (www.dirac.ac.uk). The equipment was funded by BEIS capital funding via STFC capital grants ST/K00042X/1, ST/P002293/1, ST/R002371/1 and ST/S002502/1, Durham University and STFC operations grant ST/R000832/1. DiRAC is part of the National e-Infrastructure. We also wish to acknowledge the following open source software packages used in the analysis: \textsc{Numpy} \citep{Harris_2020}, \textsc{Scipy} \citep{2020SciPy-NMeth}, \textsc{Astropy} \citep{robitaille_astropy:_2013} and \textsc{Matplotlib} \citep{Hunter:2007}. 

Most of this work was done during the coronovirus lockdown and would not have been possible without the tireless efforts of the essential workers, who did not have the safety of working from their homes. APV acknowledges the support of his PhD studentship from UK STFC DISCnet. CCL acknowledges support from the Royal Society under
grant RGF/EA/181016. PAT acknowledges support from the Science and Technology Facilities Council (grant number ST/P000525/1).

\section*{Data Availability Statement}
The photometric catalogue of the galaxies in the different regions is available at \href{https://flaresimulations.github.io/data.html}{https://flaresimulations.github.io/data.html} and the code to reproduce the plots can be found at \href{https://github.com/aswinpvijayan/flares_photometry}{https://github.com/aswinpvijayan/flares\texttt{\_}photometry}.



\bibliographystyle{mnras}
\bibliography{bib} 


\appendix

\section{Calibrating Dust Attenuation}\label{app:calibration}
\begin{figure}
	\centering
	\includegraphics[width=0.97\columnwidth]{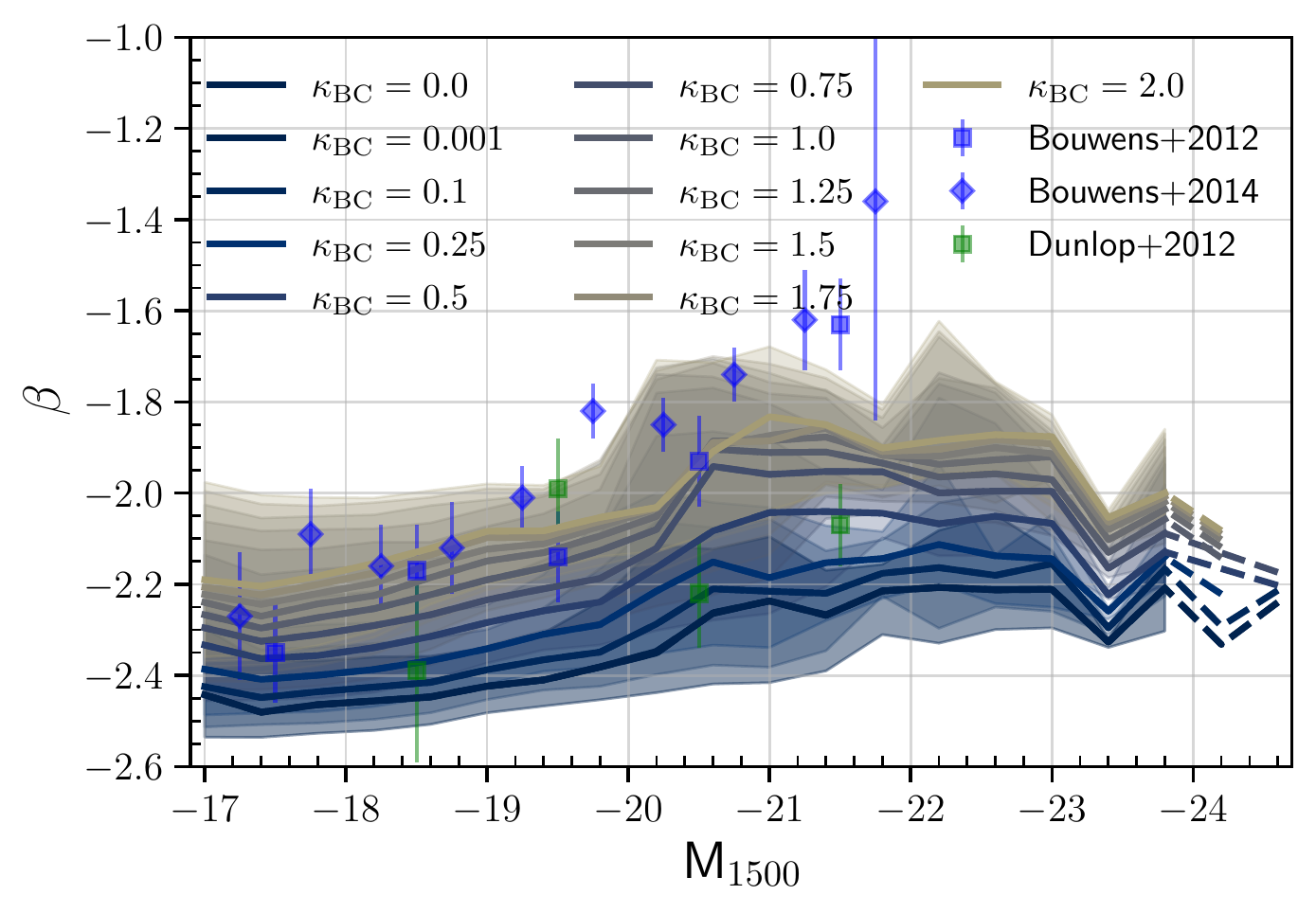}
	\caption{UV continuum slope $\beta$ for different values of $\kappa_{\textrm{BC}}$ at $z=5$. Also plotted are the observational data from \protect\cite{Dunlop2012,Bouwens2012b,Bouwens2014a}. \label{fig: beta diff kappa}}
\end{figure}
\begin{figure*}
	\centering
	\includegraphics[width=0.99\textwidth]{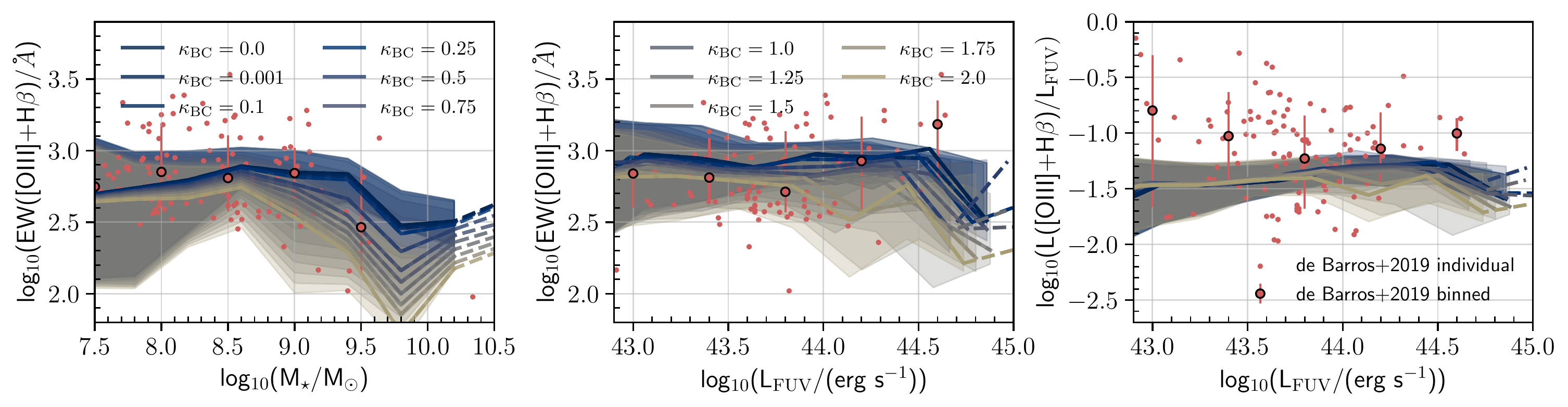}
	\caption{Same as Figure~\ref{fig: OIII_z7_8}, now showing the line luminosity and equivalent width for different values of $\kappa_{\textrm{BC}}$. The small red circles show the individual measurements from \protect\cite{deBarros19_OIIIHbeta} while the large points denote the median value in bins of stellar mass and far-UV luminosities respectively.\label{fig: line lum diff kappa}}
\end{figure*}
\begin{figure*}
	\centering
	\includegraphics[width=0.95\textwidth]{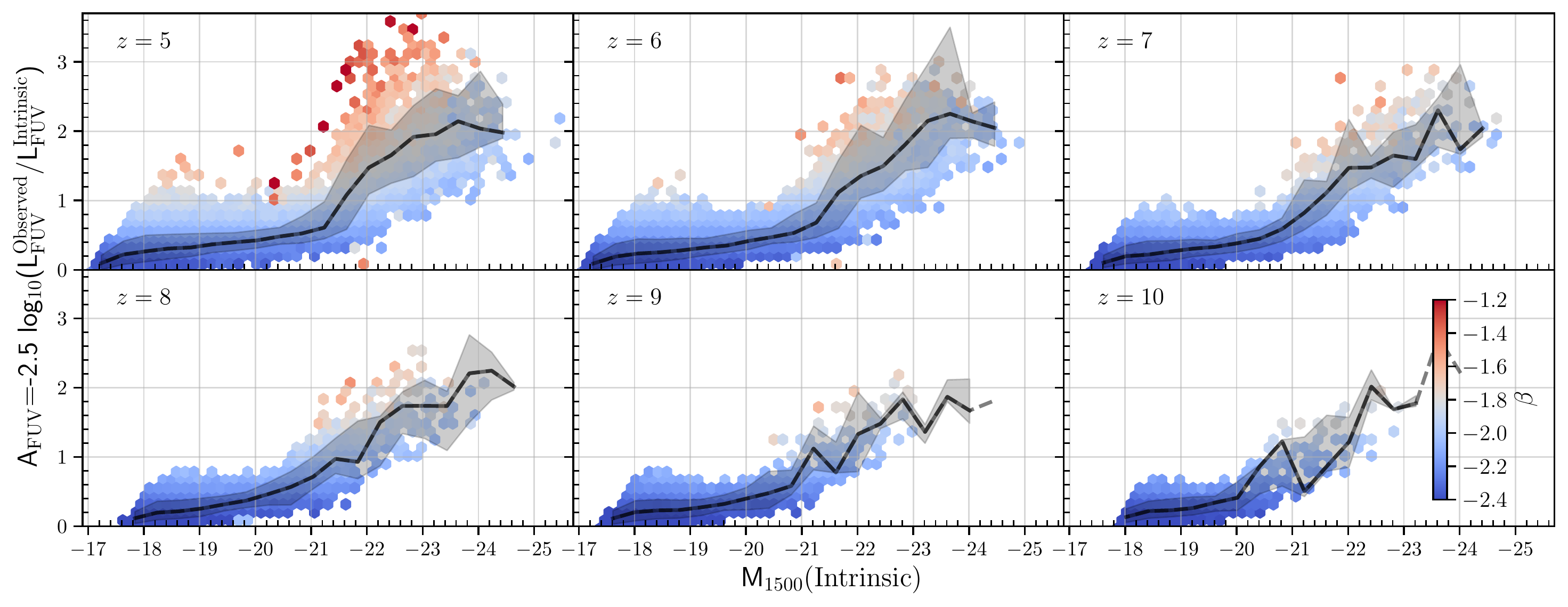}
	\caption{Same as Figure~\ref{fig: att_lfuv} and \ref{fig: att_Mstar}, now showing the attenuation as a function of intrinsic UV luminosity. \label{fig: att_lfuv_intrinsic}} 
\end{figure*}

As noted in \S\ref{sec:modelling.dust} we model the attenuation by dust on a star particle by star particle basis using the integrated line-of-sight surface density of metals as a proxy for dust attenuation. In this simple model we have a two free parameter $\kappa_{\textrm{BC}}$ and $\kappa_{\textrm{ISM}}$ which encapsulates the properties of dust such as the average grain size, shape, composition in the birth clouds and in the ISM respectively. In case of birth clouds, $\kappa_{\textrm{BC}}$ also incorporates the dust-to-metal ratio, which is assumed to scale linearly with the metallicity of the stellar particle. We calibrate these two parameters by comparing to observations of the UV LF at $z=5$ from \cite{Bouwens_2015a}, UV-continuum slope ($\beta$) at $z=5$ from \cite{Bouwens2012b,Bouwens2014a} as well as the line luminosity and the EW relation of [OIII]$\lambda$4959,5007 + H$\beta$ at $z=8$ from \cite{deBarros19_OIIIHbeta}. As explained in \S~\ref{sec:modelling.dust} we use a simple grid search to calibrate these parameters against these observations. For that purpose we generate a range of values from [0.001,2] for the parameter $\kappa_{\textrm{BC}}$. The required photometric properties\footnote{Photometric properties are generated using the code \textsc{SynthObs}: \href{https://github.com/stephenmwilkins/SynthObs}{https://github.com/stephenmwilkins/SynthObs}} are generated from $\kappa_{\textrm{ISM}}$ values in the range (0,1]. The $\kappa_{\textrm{ISM}}$ value corresponding to a given $\kappa_{\textrm{BC}}$ value is chosen to best match the UV LF from \cite{Bouwens_2015a} at $z=5$. We generate the UV LF of the \flares\, galaxies for a given ($\kappa_{\textrm{BC},i}$, $\kappa_{\textrm{ISM},j}$) pair, where `$i$' and `$j$' coresponds to a position on the grid for these parameters. The simulated and the observed UV LFs are then compared, using a chi-squared analysis to choose the best fit value of $\kappa_{\textrm{ISM}}$ for the corresponding $\kappa_{\textrm{BC},i}$. In order to select the combination of these two values that was used in this study, we compare the simulated M$_{\textrm{UV}}-\beta$ at $z=5$ against \cite{Bouwens2012b,Bouwens2014a}, shown in Figure~\ref{fig: beta diff kappa}. As can be seen from the figure, this parameter space prefers a higher value of $\kappa_{\textrm{BC}}$ for a better fit with the observational data. We tried values of $\kappa_{\textrm{BC}}>2$ and found that the median $\beta$ values have started to converge for those choices. In order to get a measure on the upper limit of $\kappa_{\textrm{BC}}$, we compare the simulated outputs of the line luminosity and the EW relation of [OIII]$\lambda$4959,5007 + H$\beta$ at $z=8$ from our range of $\kappa_{\textrm{BC}}$ choices, to the results from \cite{deBarros19_OIIIHbeta} in Figure~\ref{fig: line lum diff kappa}. As can be deduced from the figure, in this case $\kappa_{\textrm{BC}}$ prefers smaller values. In order to incorporate the impact of both these observations, we choose a value of $\kappa_{\textrm{BC}}=1$. The corresponding value of $\kappa_{\textrm{ISM}}$ is 0.0795, for this choice. Another caveat is that by fixing these values we assume there is no evolution in the general properties of the dust grains with redshift or among different galaxies.

Also presented is the relationship between the intrinsic luminosity of the galaxy and the attenuation in the far-UV in Figure~\ref{fig: att_lfuv_intrinsic}. The hexbins are coloured by their median UV continuum values ($\beta$) with the solid black line showing the weighted median and the shaded region around it representing the 84 and 16 percentiles of the data. The shape is quite similar to Figure~\ref{fig: att_lfuv} where the attenuation is plotted against the UV luminosity, and the median increases with the intrinsic luminosity and starts flattening afterwards. Also can be seen at $z=5$ is a few of the passive galaxies that have high luminosity and high $\beta$ but lower attenuation.

\section{UV LF}\label{app: lum_func}

\begin{table*}
	\centering
	\begin{tabular}[t]{cccccc}
		\hline
		M$_{1500}$ &  $\phi$\,/(cMpc$^{-3}$ Mag$^{-1}$) & M$_{1500}$ &  $\phi$\,/(cMpc$^{-3}$ Mag$^{-1}$) & M$_{1500}$ &  $\phi$\,/(cMpc$^{-3}$ Mag$^{-1}$)\\
		\hline
		\multicolumn{2}{c}{z = 5} & \multicolumn{2}{c}{z = 6} & \multicolumn{2}{c}{z = 7} \\
		\hline
		-24.286 &  (3.620$\pm$3.620)$\times 10^{-8}$ & -23.810 &  (1.473$\pm$1.473)$\times 10^{-9}$ & -23.662 &  (1.473$\pm$1.473)$\times 10^{-9}$ \\
		-23.786 &  (2.857$\pm$1.235)$\times 10^{-7}$ & -23.310 &  (3.513$\pm$3.137)$\times 10^{-6}$ & -23.162 &  (1.295$\pm$0.801)$\times 10^{-7}$ \\
		-23.286 &  (2.047$\pm$1.593)$\times 10^{-6}$ & -22.810 &  (1.008$\pm$0.311)$\times 10^{-6}$ & -22.662 &  (8.790$\pm$3.015)$\times 10^{-7}$ \\
		-22.786 & (8.674$\pm$4.616)$\times 10^{-6}$ & -22.310 &  (8.369$\pm$2.476)$\times 10^{-6}$  & -22.162 &  (4.532$\pm$2.214)$\times 10^{-6}$ \\
		-22.286 &  (2.433$\pm$0.691)$\times 10^{-5}$ & -21.810 &  (3.103$\pm$0.726)$\times 10^{-5}$ & -21.662 &  (2.326$\pm$0.632)$\times 10^{-5}$ \\
		-21.786 &  (6.266$\pm$1.186)$\times 10^{-5}$ & -21.310 &  (9.729$\pm$1.518)$\times 10^{-5}$ & -21.162 &  (5.044$\pm$1.114)$\times 10^{-5}$ \\
		-21.286 &  (1.745$\pm$0.201)$\times 10^{-4}$ & -20.810 &  (1.864$\pm$0.210)$\times 10^{-4}$ & -20.662 &  (1.168$\pm$0.164)$\times 10^{-4}$ \\
		-20.786 &  (4.484$\pm$0.339)$\times 10^{-4}$ & -20.310 &  (3.242$\pm$0.289)$\times 10^{-4}$ & -20.162 &  (1.698$\pm$0.205)$\times 10^{-4}$ \\
		-20.286 &  (7.127$\pm$0.438)$\times 10^{-4}$ & -19.810 &  (5.348$\pm$0.373)$\times 10^{-4}$ & -19.662 &  (3.745$\pm$0.320)$\times 10^{-4}$ \\
		-19.786 &  (1.043$\pm$0.053)$\times 10^{-3}$ & -19.310 &  (9.458$\pm$0.517)$\times 10^{-4}$ & -19.162 &  (6.270$\pm$0.406)$\times 10^{-4}$ \\
		-19.286 &  (1.562$\pm$0.066)$\times 10^{-3}$ & -18.810 &  (1.675$\pm$0.069)$\times 10^{-3}$ & -18.662 &  (1.381$\pm$0.062)$\times 10^{-3}$ \\
		-18.786 &  (2.634$\pm$0.087)$\times 10^{-3}$ & -18.310 &  (3.515$\pm$0.101)$\times 10^{-3}$ & -18.162 &  (3.411$\pm$0.099)$\times 10^{-3}$ \\
		-18.286 &  (4.458$\pm$0.115)$\times 10^{-3}$ & -17.810 &  (6.299$\pm$0.137)$\times 10^{-3}$ & -17.662 &  (5.898$\pm$0.133)$\times 10^{-3}$ \\
		-17.786 &  (7.703$\pm$0.152)$\times 10^{-3}$ & -17.310 &  (9.274$\pm$0.167)$\times 10^{-3}$ &   -- &  --  \\
		-17.286 &  (1.126$\pm$0.018)$\times 10^{-2}$ &   -- &  -- &  -- &   -- \\
	\end{tabular}
	\begin{tabular}[t]{cccccc}
		\hline\hline
		\multicolumn{2}{c}{z = 8} & \multicolumn{2}{c}{z = 9} & \multicolumn{2}{c}{z = 10} \\
		\hline
		-22.888 &  (2.407$\pm$2.407)$\times 10^{-8}$ & -22.662 &  (1.588$\pm$0.758)$\times 10^{-7}$ & -22.567 &  (2.407$\pm$2.407)$\times 10^{-8}$ \\
		-22.388 &  (2.429$\pm$1.545)$\times 10^{-6}$ & -22.162 &  (2.279$\pm$0.990)$\times 10^{-7}$ & -22.067 &  (4.503$\pm$3.192)$\times 10^{-8}$ \\
		-21.888 &  (1.706$\pm$0.328)$\times 10^{-6}$ & -21.662 &  (2.852$\pm$1.624)$\times 10^{-6}$ & -21.567 &  (2.075$\pm$1.538)$\times 10^{-7}$ \\
		-21.388 &  (1.675$\pm$0.484)$\times 10^{-5}$ & -21.162 &  (1.098$\pm$0.414)$\times 10^{-5}$ & -21.067 &  (1.130$\pm$0.526)$\times 10^{-5}$ \\
		-20.888 &  (4.410$\pm$1.002)$\times 10^{-5}$ & -20.662 &  (3.000$\pm$0.880)$\times 10^{-5}$ & -20.567 &  (6.563$\pm$1.951)$\times 10^{-6}$ \\
		-20.388 &  (7.125$\pm$1.379)$\times 10^{-5}$ & -20.162 &  (4.470$\pm$1.041)$\times 10^{-5}$ & -20.067 &  (1.251$\pm$0.423)$\times 10^{-5}$ \\
		-19.888 &  (1.186$\pm$0.178)$\times 10^{-4}$ & -19.662 &  (8.275$\pm$1.420)$\times 10^{-5}$ & -19.567 &  (5.984$\pm$1.237)$\times 10^{-5}$ \\
		-19.388 &  (2.473$\pm$0.254)$\times 10^{-4}$ & -19.162 &  (2.236$\pm$0.244)$\times 10^{-4}$ & -19.067 &  (1.764$\pm$0.214)$\times 10^{-4}$ \\
		-18.888 &  (6.183$\pm$0.409)$\times 10^{-4}$ & -18.662 &  (7.084$\pm$0.444)$\times 10^{-4}$ & -18.567 &  (5.418$\pm$0.387)$\times 10^{-4}$ \\
		-18.388 &  (1.732$\pm$0.070)$\times 10^{-3}$ & -18.162 &  (1.844$\pm$0.073)$\times 10^{-3}$ & -18.067 & (1.473$\pm$0.064)$\times 10^{-3}$ \\
		-17.888 &  (3.329$\pm$0.098)$\times 10^{-3}$ & -17.662 &  (3.027$\pm$0.094)$\times 10^{-3}$ &   -- &  -- \\
		\hline
	\end{tabular}
	\caption{Binned UV LF values for the \flares\, galaxies. Also quoted is the weighted 1-$\sigma$ Poisson uncertainty for the number density within each luminosity bin. \label{tab: LF values}}
\end{table*}
\begin{table*}
	\centering
	\begin{tabular}[t]{cccccc}
		\hline
		z &  M$^{*}$/Mag &  log$_{10}$($\phi^{*}$\,/(Mpc$^{-3}$ Mag$^{-1}$)) &  $\alpha$ & $\beta$ & $\Delta$BIC\\
		\hline\\
		& -21.844$^{+0.041}_{-0.042}$ & -3.662$^{+0.025}_{-0.025}$ & -1.984$^{+0.006}_{-0.006}$ & --\\5&&&&& 66.440\\
		& -21.699$^{+0.035}_{-0.042}$ & -3.766$^{+0.022}_{-0.026}$ & -2.033$^{+0.006}_{-0.007}$ & -4.406$^{+0.139}_{-0.130}$\\\\
		& -21.666$^{+0.039}_{-0.040}$ & -3.946$^{+0.027}_{-0.027}$ & -2.151$^{+0.007}_{-0.007}$ & --\\6&&&&& 37.711\\
		& -21.819$^{+0.036}_{-0.034}$ & -4.224$^{+0.023}_{-0.024}$ & -2.226$^{+0.006}_{-0.006}$ & -5.232$^{+0.097}_{-0.050}$\\\\
		& -22.226$^{+0.088}_{-0.092}$ & -4.859$^{+0.067}_{-0.070}$ & -2.483$^{+0.012}_{-0.012}$ & --\\7&&&&& 48.069\\
		& -22.104$^{+0.076}_{-0.061}$ & -4.934$^{+0.055}_{-0.046}$ & -2.522$^{+0.011}_{-0.010}$ & -5.235$^{+0.125}_{-0.048}$\\\\
		& -22.082$^{+0.157}_{-0.135}$ & -5.307$^{+0.134}_{-0.115}$ & -2.732$^{+0.019}_{-0.016}$ & --\\8&&&&& 11.431\\
		& -21.841$^{+0.102}_{-0.101}$ & -5.281$^{+0.085}_{-0.086}$ & -2.771$^{+0.016}_{-0.015}$ & -5.179$^{+0.216}_{-0.092}$\\\\
		& -21.224$^{+0.174}_{-0.157}$ & -4.838$^{+0.147}_{-0.135}$ & -2.702$^{+0.024}_{-0.022}$ & --\\9&&&&& 67.436\\
		& -19.023$^{+0.018}_{-0.044}$ & -3.148$^{+0.016}_{-0.036}$ & -2.304$^{+0.029}_{-0.033}$ & -3.730$^{+0.072}_{-0.080}$\\\\
		& -20.453$^{+0.284}_{-0.242}$ & -4.768$^{+0.320}_{-0.255}$ & -3.136$^{+0.077}_{-0.046}$ & --\\10&&&&& 3.410\\
		& -19.491$^{+0.359}_{-0.632}$ & -3.900$^{+0.362}_{-0.663}$ & -3.025$^{+0.127}_{-0.142}$ & -4.136$^{+0.193}_{-0.239}$\\\\
		\hline
	\end{tabular}
	\caption{Best-fitting Schechter (first row corresponding to the redshift) and double power-law (second row corresponding to the redshift) function parameter values for the observed UV LF. The quoted error bars show the $16^{\mathrm{th}}-84^{\mathrm{th}}$ percentile uncertainty obtained from the fit posteriors. We also provide the difference of the Bayesian Information Criterion ($\Delta$BIC) value of the best-fitting parameters of the double power-law from the Schechter function. \label{tab: fit params}}
\end{table*}
For deriving the Schechter and double power-law fit parameters for the UV LF, we calculate the likelihood that the number of observed galaxies in a given magnitude bin is equal to that for an assumed value of the function parameters. This calculation is performed in bins of separation $\Delta M = 0.5$ mag, ranging from from our completeness limit at the faint-end to enclose all our galaxies above this limit. Bins containing less than 5 galaxies were not considered while fitting. The bin centre and the number density of galaxies per magnitude is provided in Table~\ref{tab: LF values}. We use the code \textsf{FitDF}\footnote{\href{https://github.com/flaresimulations/fitDF}{https://github.com/flaresimulations/fitDF}} a Python module for fitting arbitrary distribution functions. \textsf{FitDF} uses \textsf{emcee}, a Python implementation of the affine-invariant ensemble sampler for Markov chain Monte Carlo (MCMC) described in \cite{DFM2013}. The likelihood function is modelled as a Gaussian distribution of the following form
\begin{align}\label{eq: likelihood}
\mathrm{ln}(\mathcal{L}) = - \frac{1}{2} \,  \sum_{i} \,\left[ \frac{ (n_{i,\mathrm{obs}} - n_{i,\mathrm{exp}})^2 }{\sigma_{i}^{2}} \, + \mathrm{log}(\sigma_{i}^{2}) \right] \;\;,
\end{align}
where the subscript $i$ represents the bin of the property being measured, $n_{i,\mathrm{obs}}$ is the number density of galaxies using the composite number density, $n_{i,\mathrm{exp}}$ is the expected number density from the functional form being used (Schechter or double power-law), and $\sigma_i$ is the error estimate. Using this form, $\sigma$ can be explicitly provided by the expression, $\sigma_{i} = n_{i,\mathrm{obs}} / \sqrt{N_{i,\mathrm{obs}}}$, where $N_{i,\mathrm{obs}}$ is the number counts in bin $i$ from the re-simulations. We use flat uniform priors for the parameters in the functional forms. In the case of the double power-law form, to constrain the parameters, $\beta$ was restricted to a lower limit of $-5.3$ and M$^{*}$ to an upper limit of $-19$. 

For determining which functional form is better suited at different redshifts we calculated the Bayesian Information Criterion (BIC) value for the best-fit parameters. BIC is a criterion for model selection among a finite set of models, defined as follows:
\begin{equation}\label{eq: BIC}
\mathrm{BIC} = -2\mathrm{ln}(\mathcal{L})+\mathrm{k ln(N)}\:,
\end{equation} 
where $\mathcal{L}$ is the likelihood of the fit function as expressed in Equation~\ref{eq: likelihood}, k is the number of free
parameters, and N is the number of data points going into the fitting. When performing fitting it is possible to increase the likelihood by adding more parameters, but can lead to overfitting. BIC resolves this by implementing a penalty term for the number of parameters in the model; the model with a lower BIC is preferred. A difference of $\ge 20$ in the BIC value is usually taken to be a very strong preference for the model with a lower values. The difference of the BIC values, $\Delta\textrm{BIC}$ of the double power-law from the Schechter functional form is shown in Table~\ref{tab: fit params}.

\section{Other extinction curves}\label{app:extinction_curves}

\begin{figure*}
	\centering
	\includegraphics[width=0.95\columnwidth]{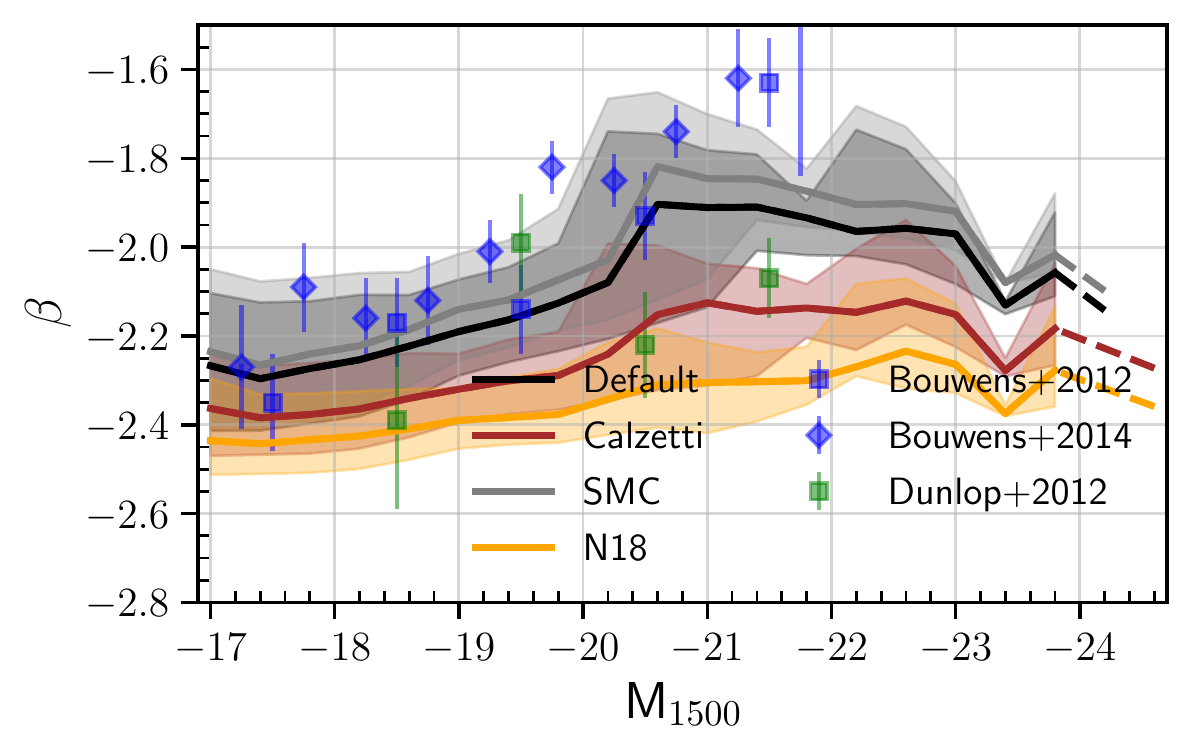}
	\includegraphics[width=1.08\columnwidth]{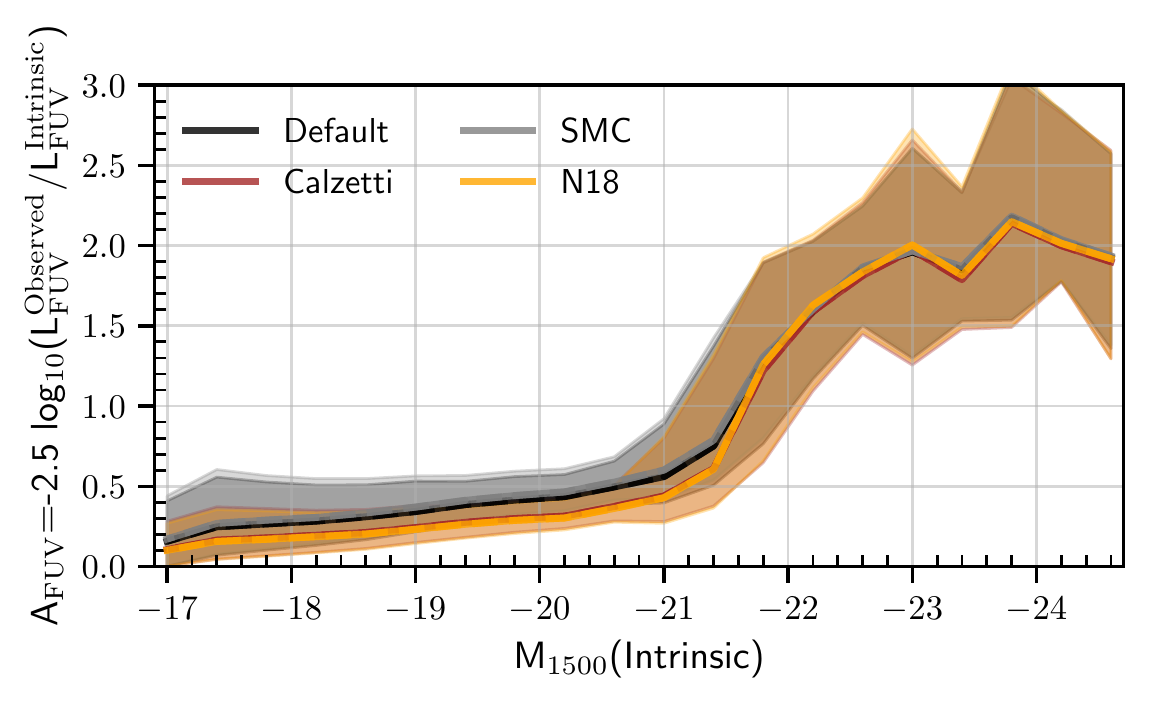}
	\caption{Left: Same as Figure~\ref{fig: beta diff kappa}, now showing $\beta$ values for different extinction curves. Right: Attenuation in far-UV for different extinction curves at $z=5$. Solid lines denotes the weighted median of the sample.\label{fig: ext curves beta att}}
\end{figure*}

\begin{figure*}
	\centering
	\includegraphics[width=0.99\textwidth]{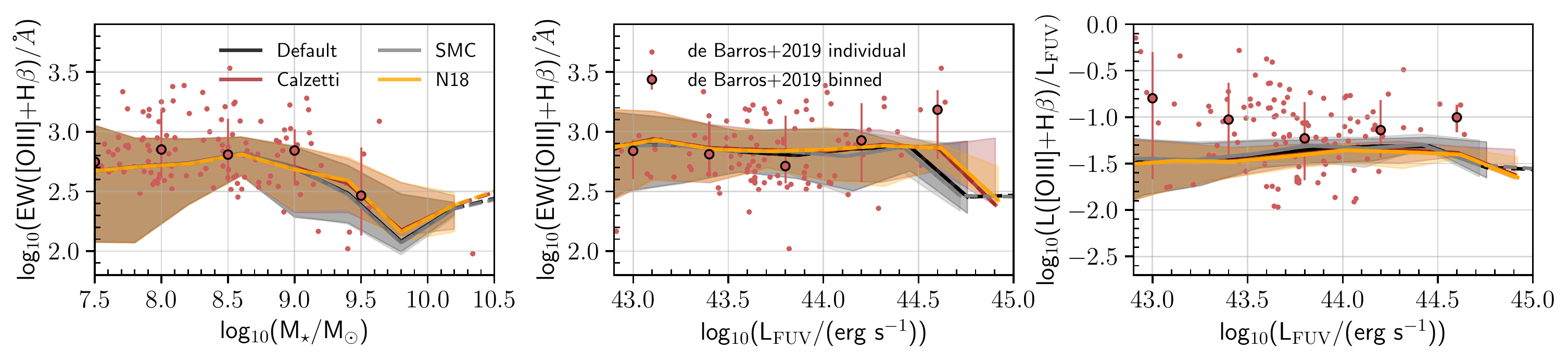}
	\caption{Same as Figure~\ref{fig: OIII_z7_8}, now showing the line luminosity and equivalent widths for for different extinction curves. \label{fig: ext curves EW}}
\end{figure*}
There has not been any consensus across observational or theoretical studies on the exact nature of the extinction curve in galaxies, since it is closely tied to the properties of the dust grains in galaxies. And this can be inferred better by probing the galaxy SED, and studies have suggested that using a single extinction curve for every galaxy might not be right. In our study we implement a simple extinction curve that is inversely proportional to the wavelength. In this section we will explore how some of the observables presented before changes depending on the chosen extinction curve, namely the Calzetti \citep{Calzetti2000}, Small Magellanic Cloud \citep[SMC, ][]{Pei1992} and the curve used in \cite[][N18 from now on]{Narayanan2018}. 

For this analysis we keep the value of $\kappa_{\textrm{BC}}$ from our default model curve, \ie\, $\kappa_{\textrm{BC}}=1.0$. We then use the method described in Appendix~\ref{app:calibration} to get $\kappa_{\textrm{ISM}}$, obtaining the values of 0.175, 0.0691 and 0.22 for the Calzetti, SMC and N18 curves respectively.

In the left panel of Figure~\ref{fig: ext curves beta att} we present the effect of using different attenuation curves on the UV continuum slope, $\beta$. It can be seen the SMC curve has a higher median for the UV continuum slope, compared to the default model, a consequence of the SMC curve being steeper than our default value. While for the case of the Calzetti and N18 curves the former has a higher normalisation compared to the latter. We also tried increasing the value of $\kappa_{\textrm{BC}}$ for the Calzetti and N18 curves to steepen the relation. We find that the match to the steepness of the observations is difficult to obtain from these curves, implying the \flares\, galaxies prefer a steeper extinction curve similar to the SMC to reproduce the UV continuum observations.

In the right panel of Figure~\ref{fig: ext curves beta att} we present the effect of using different attenuation curves on the attenuation in the far-UV. There is no observed difference in the attenuation in the FUV for any of the curves except at intrinsic M$_{1500}\gtrapprox-21.5$ where the Calzetti and N18 curves produce on average lower attenuation. From our discussion before it is quite clear that despite this the underlying properties vary differently on using these different extinction curves. 

In Figure~\ref{fig: ext curves EW} we present the effect of using different attenuation curves on the line luminosity and equivalent width relationship of the [OIII]$\lambda$4959,5007 doublet. As can be seen all the curves trace the same space in all the sub-figures. Any minute difference seen happens at higher stellar mass/far-UV luminosity, with the default and SMC curve tracing a slightly lower median than the others.



\bsp	
\label{lastpage}
\end{document}